\documentclass[aps,prl,twocolumn,floatfix,
letterpaper,superscriptaddress]{revtex4}
\usepackage{graphicx}\usepackage{natbib}\usepackage{amsmath}
\newcommand{\kms}{\ensuremath{{\rm km\,s}^{-1}}}

\newcommand{\yr}{\ensuremath{\rm yr}}

\begin{document}

\title{Young Pulsars and the Galactic Center GeV Gamma-ray Excess}

\author{Ryan M.\ O'Leary}
\affiliation{JILA, 440 UCB, University of Colorado, Boulder, CO 80309-0440, USA}
\author{Matthew D.\ Kistler}
\affiliation{Kavli Institute for Particle Astrophysics and Cosmology, Stanford University,
SLAC National Accelerator Laboratory, Menlo Park, CA 94025, USA}
\author{Matthew Kerr}
\affiliation{CSIRO Astronomy and Space Science, Australia Telescope National Facility, PO Box 76, Epping NSW 171, Australia}
\author{Jason Dexter}
\affiliation{Max Planck Institute for Extraterrestrial Physics, Giessenbachstr. 1, 85748 Garching, Germany}

\date{April 9, 2015}

\begin{abstract}
Studies of {\em Fermi} data indicate an excess of GeV gamma rays around the Galactic center (GC), possibly due to dark matter.  We show that young gamma-ray pulsars can yield a similar signal.  First, a high concentration of GC supernovae naturally leads to a population of kicked pulsars symmetric about the GC.  Second, while very-young pulsars with soft spectra reside near the Galactic plane, pulsars with spectra that have hardened with age accumulate at larger angles. This combination, including unresolved foreground pulsars, traces the morphology and spectrum of the Excess.

\end{abstract}

% 98.70.Rz     gamma ray sources; gamma ray bursts
% 98.70.-f 	   Unidentified sources of radiation outside the Solar System
%,showpacs
 %95.85.Ry, 98.70.Rz, 98.70.-f}
\maketitle

\textbf{Introduction.}---
The {\em Fermi} Large Area Telescope \cite{2009ApJ...697.1071A} has transformed GeV gamma-ray astrophysics.  Pulsar physics in particular has experienced an enormous advance \cite{2014ARA&A..52..211C}.  The impact is much greater than even the impressive increase in the number of gamma-ray pulsars, from 7 pre-{\em Fermi} to $>$170 now \cite{2013pulsarcatalog,2015arXiv150203251L}, alone suggests.  GeV detections of old, recycled millisecond pulsars (MSPs) went from zero to $\sim\,$60 in under six years.  The $>\,$100 detections of young pulsars ($\tau \lesssim$~Myr) imply ubiquitous GeV emission \cite{2011ApJ...727..123W,2012A&A...545A..42P}, with the radio-quiet population now at $\sim\,$40 from a lone prototypical member, Geminga \cite{1992Natur.357..222H,1992Natur.357..306B}.

{\em Fermi} could also fulfill the long-standing hope of detecting gamma rays from the annihilation or decay of dark matter \cite{1996PhR...267..195J,2004JCAP...07..008G,2005PhR...405..279B,2013PhR...531....1S,2014PhRvD..89d2001A,2015arXiv150302641F}, which comprises $\approx 84\,\%$ of all matter \cite{1933AcHPh...6..110Z,2015arXiv150201589P}.  A number of groups \cite{2009arXiv0910.2998G,2011PhLB..697..412H,2011PhRvD..84l3005H,2012PhRvD..86h3511A,2013PDU.....2..118H,2013PhRvD..88h3521G,2014PhRvD..89f3515M,2014PhRvD..90b3526A,2014daylanetal,2014arXiv1406.6948Z,2014arXiv1409.0042C,2014caloreetal,MurgiaTalk} have used {\em Fermi} data to search for a signal originating from around the Galactic center (GC), where this flux should be largest.

The exciting recent development from these studies is the measurement of an extended excess of GeV gamma rays above model predictions peaking from 1--3~GeV.  The Excess spectrum is reasonably fit by the annihilation of a $30-60$~GeV dark matter particle \cite[e.g.,][]{2011PhRvD..84l3005H,2012PhRvD..86h3511A,2013PDU.....2..118H,2013PhRvD..88h3521G,2014PhRvD..89f3515M,2014PhRvD..90b3526A,2014daylanetal,2014arXiv1406.6948Z,2014arXiv1409.0042C,2014caloreetal,MurgiaTalk,2014PhRvD..90c5004A,2014PhRvD..90e5002I,2014PhRvD..90e5021I,2014NuPhB.888..154K,2014JCAP...09..013K,2014PhRvD..90b3531B,2014PhRvD..90g5011C,2014arXiv1410.3818F,2014arXiv1411.2592A,2014PhRvD..90j3513M,2014arXiv1412.1485L,2015PhRvD..91b3505B,2015JHEP...02..057C,2015arXiv150206000B,2015arXiv150207173G,2015arXiv150301773E}, though the angular intensity implies a density profile steeper than often extrapolated from simulations.

The great challenge is to determine whether this signal indeed arises from dark matter or is due to an unaccounted-for source of gamma rays.  One suggestion is a contribution from MSPs \cite[][]{2012PhRvD..86h3511A,2013MNRAS.436.2461M,2014JHEAp...3....1Y,2014ApJ...796...14C}, which have a similar spectral shape as the Excess in gamma rays and are expected to exist there at some level \cite{fauchergiguereloeb2011}.  It has been argued that MSPs cannot match the Excess \cite{2013PhRvD..88h3009H,2014arXiv1407.5583C} and uncertainties in MSP formation make definite predictions difficult.

%%%%%%%%%%%%%%%%
%%%%%%%%%%%%%%%%
\begin{figure}[b!]
\includegraphics[width=\columnwidth,clip=true]{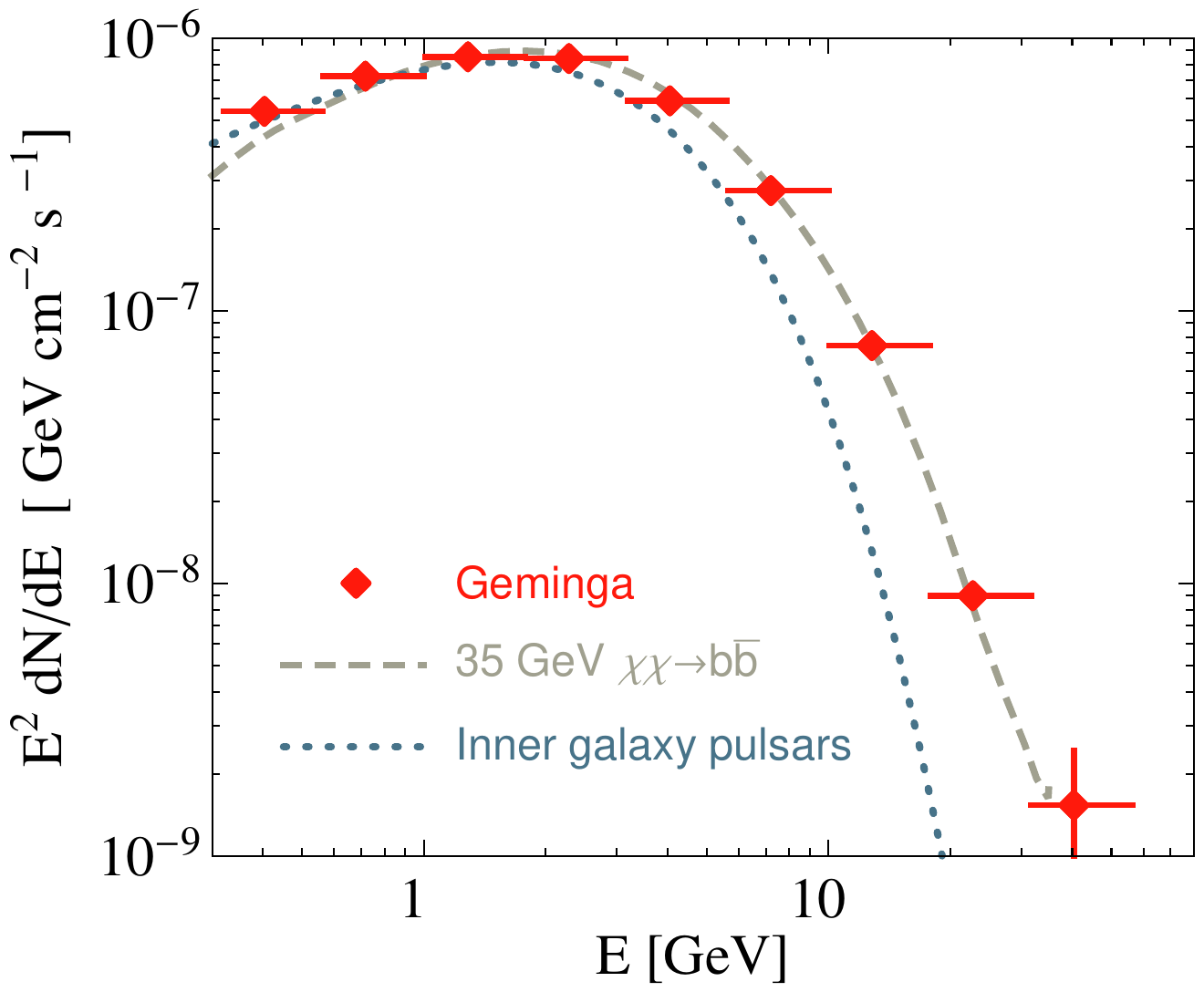}\vspace{-0.4cm}
\caption{A comparison of phase-averaged {\it Fermi} gamma-ray data of Geminga \cite{2015ApJ...800...61A} to our spectral shape from pulsars within $5^\circ$ of the GC and $>2^\circ$ from the disk (see text), which resemble that from $35\,$GeV dark matter particles annihilating to $b\bar{b}$ as proposed to explain the GC Excess \cite{2014daylanetal}.  \label{fig:geminga}}
\end{figure}
%%%%%%%%%%%%%%%%
%%%%%%%%%%%%%%%%

We show that young pulsars from the GC itself produce a diffuse GeV flux that has been widely underestimated.
Such pulsars arise from core-collapse supernova explosions of short-lived massive stars.  Their birthplaces thus tend to trace star formation.  The GC is the most concentrated star forming region in the Milky Way (MW); the inner $\sim 200$~pc central molecular zone (CMZ) accounts for $5 - 10\%$ of the current Galactic star formation rate \cite[e.g.,][]{1996ARA&A..34..645M,genzeletal2010,2012ARA&A..50..531K,2013MNRAS.429..987L}.  Further, the CMZ contains an estimated $\sim 13\%$ of all MW Wolf rayet stars \cite{2015MNRAS.447.2322R,2007ARA&A..45..177C,2010ApJ...725..188M,2012MNRAS.425..884D,2014arXiv1403.0975K}; these evolved, $\gtrsim\,$$25\,M_\odot$ stars are near explosion and may remain from an even more intense recent period of star formation \cite[e.g.,][]{2014MNRAS.440.3370K}.
This implies a substantial, ongoing production of pulsars, with birth kicks leading to a continuous, symmetrical distribution centered on the GC.

Fig.~\ref{fig:geminga} displays a second piece to this puzzle.  Here we 
compare the gamma-ray spectrum measured from Geminga, the best characterized of radio-quiet pulsars, to that expected from 35~GeV dark matter annihilating to a $b$-quark pair \cite{2011JCAP...03..051C} as proposed for the Excess. The striking similarity to the Excess is a consequence of the hardening of gamma-ray pulsar spectra with decreasing spin-down power ($\dot{E}$) as seen in {\em Fermi} data \cite{2013pulsarcatalog,2015A&A...575A...3P}.  

Very young, high-$\dot{E}$ pulsars with soft spectra then reside within the Galactic plane, where they are masked or subtracted as point sources and do not contribute to the observed Excess. On the other hand, pulsars old enough to reach large angles have hard spectra (Geminga has a characteristic age of $\sim\,$$3\!\times\!10^5\,$yr).  {\em Fermi} has also shown that the gamma-ray luminosity, $L_\gamma \propto \dot{E}^{1/2}$, declines much more slowly than spin-down power, while gamma-ray efficiency, $L_\gamma/\dot{E}$, rises at low $\dot{E}$ \cite{2013pulsarcatalog}.  Since an additional $20\%$ of Galactic star formation occurs within a projected radius of $15^\circ$ of the GC \cite{2004A&A...422..545Y}, there should be unresolved high-latitude disk pulsars that also contribute to the Excess.  Indeed, $\sim 15\%$ of all radio pulsars are seen in this region \cite{manchesteretal2005}.

Our conservative estimates suggest the population of unresolved young pulsars may constitute a significant fraction, if not all, of the excess GeV gamma rays.
Thanks to intensive {\em Fermi} and radio studies, we
are able to forward model the emission of young pulsars. This is in
contrast to modeling MSPs, which have complicated formation channels,
or ancient cosmic-ray bursts
\cite{2014PhRvD..90b3015C,2014JCAP...10..052P} or dark matter which
have a large range of free parameters.  We do not include any MSP or other non-young-pulsar contribution in this work.  We discuss a few of the many
implications here and in a companion paper \cite{2015toappear}.

\textbf{The GC pulsar factory.}---
We use a suite of Monte Carlo simulations to forward model the present day
 distribution of ordinary young pulsars throughout the GC and Galactic disk.  Our
 methods largely follow Refs.~\cite{fauchergiguerekaspi2006} and \cite{2011ApJ...727..123W}, who successfully 
 reproduced the observed distribution of nearby radio and gamma-ray pulsars, respectively. We extend both by including the CMZ contribution (avoiding here the peculiar central parsec \cite{dexter2014}).  

To generate our
 simulated gamma-ray pulsar population, we select the pulsars' ages
 randomly assuming a constant birth rate of $2.1\times
 10^{-2}\,\yr^{-1}$ in the disk
 \cite{fauchergiguerekaspi2006,2011MNRAS.412.1473L}, formed  on circular orbits within the MW
 spiral arms \cite{fauchergiguerekaspi2006}.

The GC rate depends on the number of massive stars formed over the previous $\sim 30$~Myr, due to their finite ages \cite{2013A&A...558A.131G}, while the present star formation rate may be near a minimum if episodic on $\sim 10\,$Myr periods \cite{2014MNRAS.440.3370K}.  Over this time their positions will be symmetrized by cluster dispersion and differential rotation.  For simplicity, we assume a constant birth rate of $1.5\times 10^{-3}\,\yr^{-1}$ in the CMZ, $\approx 7\%$ of the Galactic rate.  These pulsars are formed from a spherical isothermal  distribution ($n \propto r^{-2}$), between 20 and 200\,pc of Sgr A* -- the region of most GC star formation.

 We kick each pulsar by selecting a random kick velocity from a three-dimensional normal distribution with a mean velocity of $\approx 408\,\kms$ \cite{2005MNRAS.360..974H}.  Each pulsar orbit is integrated over its lifetime using
 {\sc galpy} \footnote{Available at
   {\protect\url{https://github.com/jobovy/galpy}}} with the
 \texttt{MWPotential2014} model for the Galactic potential \citep{bovy2015}.

The pulsar gamma-ray luminosity depends on its total spin-down power, which in our models is determined by the surface magnetic field, initial spin period, and age.   We assign each pulsar a constant surface $B$ from a log-normal distribution
with mean $\left<\log_{10}{B}\right> = 12.65$ and $\sigma_{\log{B}} = 0.45$.  The variance in the magnetic fields is roughly between those used by Refs.~\cite{fauchergiguerekaspi2006} \& \cite{2011ApJ...727..123W}.
The pulsars are born with an initial spin
period, $P_0$ following \cite{2011ApJ...727..123W}. 
Therefore the spin-down luminosity of the pulsar $\dot{E} = 4 \pi^2 I \dot{P} P^{-3}$, where $I \equiv  1\times 10^{45}\,$g$\,$cm$^2$ is the pulsar moment of inertia, $\dot{P} = 8 \pi^2 R^6 B^2/ (3 I c^3 P)$, $R\equiv 10^6\,$cm is the radius of the pulsar, $c$ is the speed of light, and $P = (P_0^2+16 \pi^2 R^6 B^2 t / (3 I c^3))^{1/2}$.

%%%%%%%%%%%%%%%%
%%%%%%%%%%%%%%%%
\begin{figure}[t!]
\hspace*{-0.45cm}
\includegraphics[width=1.05 \columnwidth,clip=true]{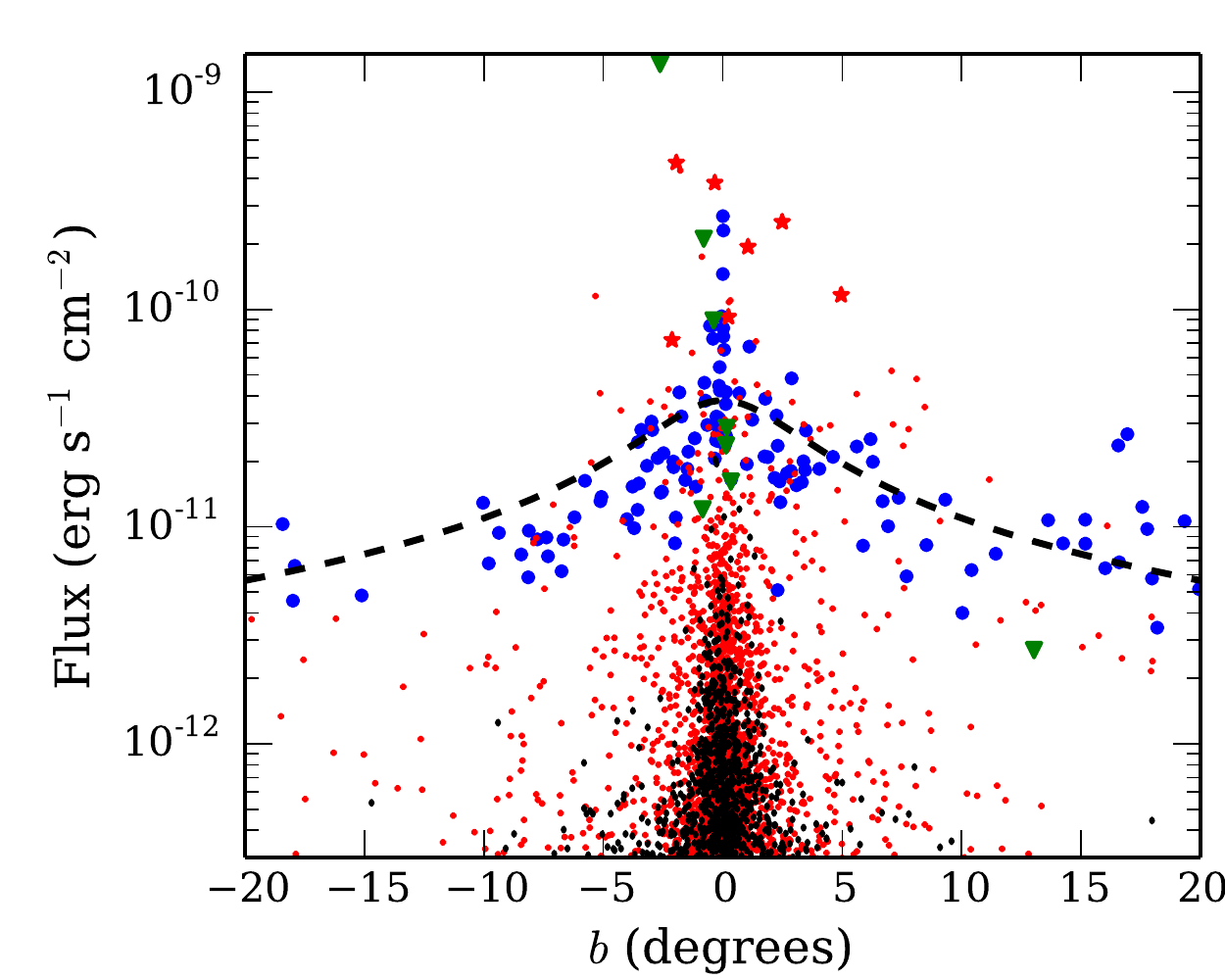}
\caption{Fluxes of simulated pulsars within Galactic longitude $|l|<20^\circ$ and latitude $|b|<20^\circ$
from one realization of our fiducial model, separated into those from the GC (black {\it diamonds}) and disk (small red {\it dots}).
 These are compared to the population of {\it Fermi} 3FGL point sources (blue {\it circles}), gamma-ray discovered pulsars (red {\it stars}),  and radio discovered pulsars (green {\it triangles}) \citep{2013pulsarcatalog,2014A&A...570A..44H,2015arXiv150203251L}. The dashed line shows the 50\% completeness limit for 2FGL \cite[see our Eq.~(\ref{eq:ps});][]{2012ApJS..199...31N}.
  \label{fig:pointsource}}
\end{figure}
%%%%%%%%%%%%%%%%
%%%%%%%%%%%%%%%%

\textbf{Gamma rays in excess.}---
{\it Fermi} observations of young pulsars show that their
sky-averaged gamma-ray luminosity follows a trend that agrees with
the theoretically expected $L_{\gamma} \propto \dot{E}^{1/2}$
\citep[e.g.,][]{2013pulsarcatalog,2011ApJ...727..123W}, although with notable scatter.  
We normalize $L_{\gamma}$ for each pulsar
as
\begin{equation}
\label{eq:lumfunc}
L_\gamma =  C \times \left(\frac{\dot{E}}{10^{33}\,{\rm erg/s}}\right)^{1/2} \times 10^{33}\,{\rm erg/s}.
\end{equation}
The heuristic value of $C = 1.0$
\cite{2011ApJ...727..123W,2013pulsarcatalog} underestimates the
luminosity of Geminga by a factor of 5, and the mean and median
gamma-ray luminosity of ordinary pulsars found in
\cite{2013pulsarcatalog} by a factor of $\approx 3$.  For a given
pulsar, models predict different beaming corrections to the observed
flux dependent upon the angle to the observer and magnetic field
alignment \cite{2009ApJ...695.1289W}.  Using the beaming-corrected
relations of \cite{2015A&A...575A...3P} modeled from many individual
light curves, we find that the mean luminosity is $C=2.2$, after
removing the top and bottom 5\% of the sample.  Here, we are mostly
concerned with a population-averaged luminosity density and take a
fiducial value of $C = 1.3$, which produces one-third of the
unassociated plane sources in the {\it Fermi} 3FGL catalog
\cite{2015arXiv150102003T}.  For simplicity, we assume that gamma-ray
emission turns off at $\dot{E} \lesssim 10^{33.5}\,$erg~s$^{-1}$,
higher than the outer gap death line of
\cite{1996A&AS..120C..49A,2011ApJ...736..127W}.

In Fig.~\ref{fig:pointsource}, we show the gamma-ray fluxes of a
single realization of the GC and Galactic disk pulsar populations as a
function of Galactic latitude, $b$. Also plotted are the fluxes of
3FGL point sources \cite{2015arXiv150102003T} and gamma-ray
 pulsars \cite{2013pulsarcatalog,2014A&A...570A..44H,2015arXiv150203251L} within the same region. In
our model, no pulsars formed at the GC are bright enough to be
detected as a distinct point source by {\em Fermi}.  Indeed, only
pulsars from the Galactic disk cross the 50\% point
source completeness limit we later use for subtraction.

This same region contains seven gamma-ray-discovered pulsars and our
simulations typically contain a comparable number. We also find that
the results of our simulations are compatible with the entire Galactic
pulsar population.  Like previous analyses \cite{2011ApJ...727..123W,2012A&A...545A..42P},
we produce too few of the most luminous pulsars. Increasing $C$ helps to alleviate some of this tension (see \cite{2015toappear} for a discussion of this and pulsar beaming).
Fig.~\ref{fig:pointsource} also shows that, on average, the highest $L_\gamma$
GC pulsars are the closest to the GC, a natural consequence of
spin-down. In our model, pulsar gamma-ray luminosity decreases with time as $t^{-1}$.

%%%%%%%%%%%%%%%%
%%%%%%%%%%%%%%%%
\begin{figure}[t!]
  \includegraphics[width=\columnwidth,clip=true]{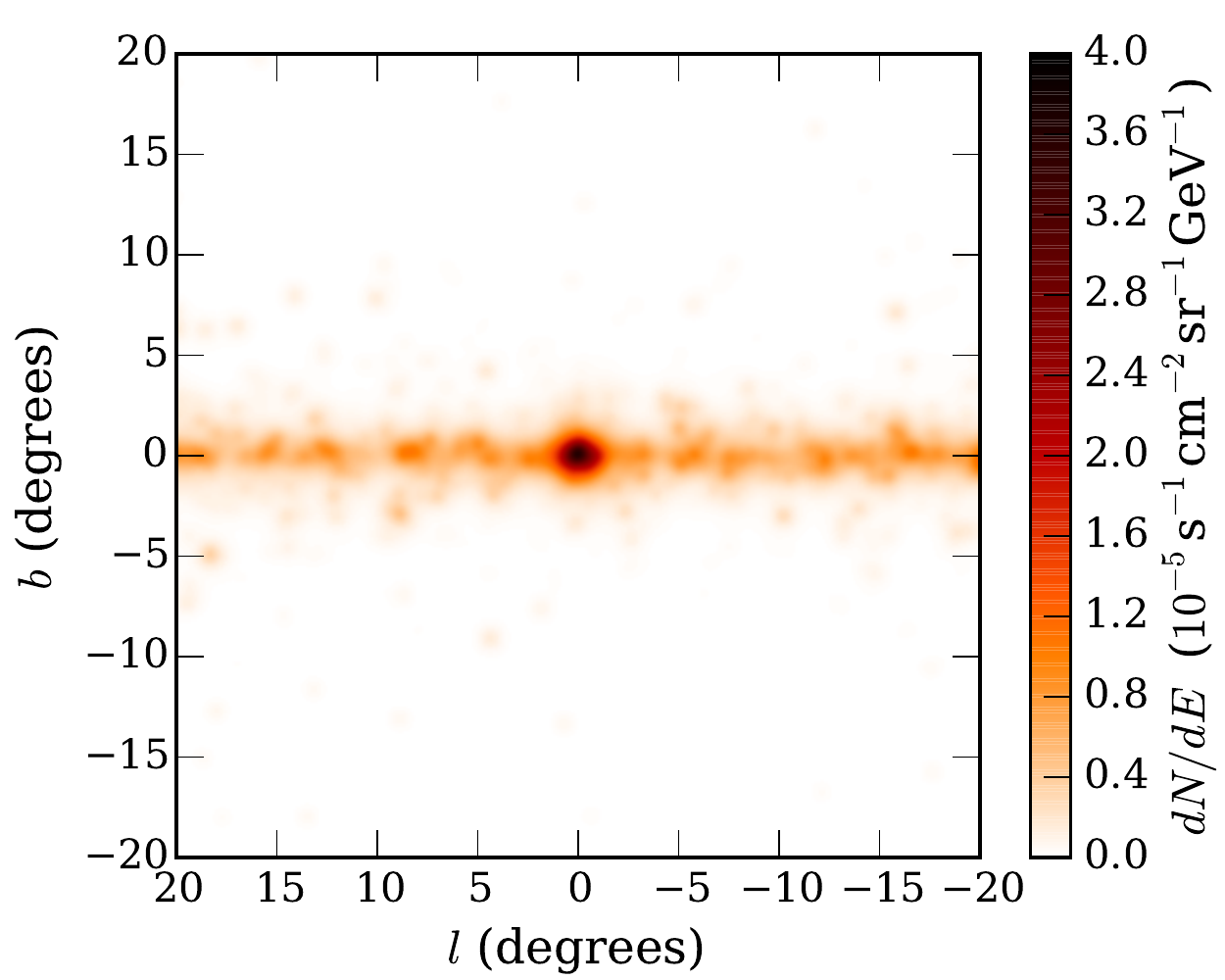}
\caption{The spatial distribution of 2 GeV gamma rays from {\it unresolved} young pulsars in one realization.  Note that we do not include detector noise or other Galactic emission here. \label{fig:cmz}}
\end{figure}
%%%%%%%%%%%%%%%%
%%%%%%%%%%%%%%%%

%%%%%%%%%%%%%%%%
%%%%%%%%%%%%%%%%
\begin{figure*}[t!]
\includegraphics[width=\textwidth,clip=true]{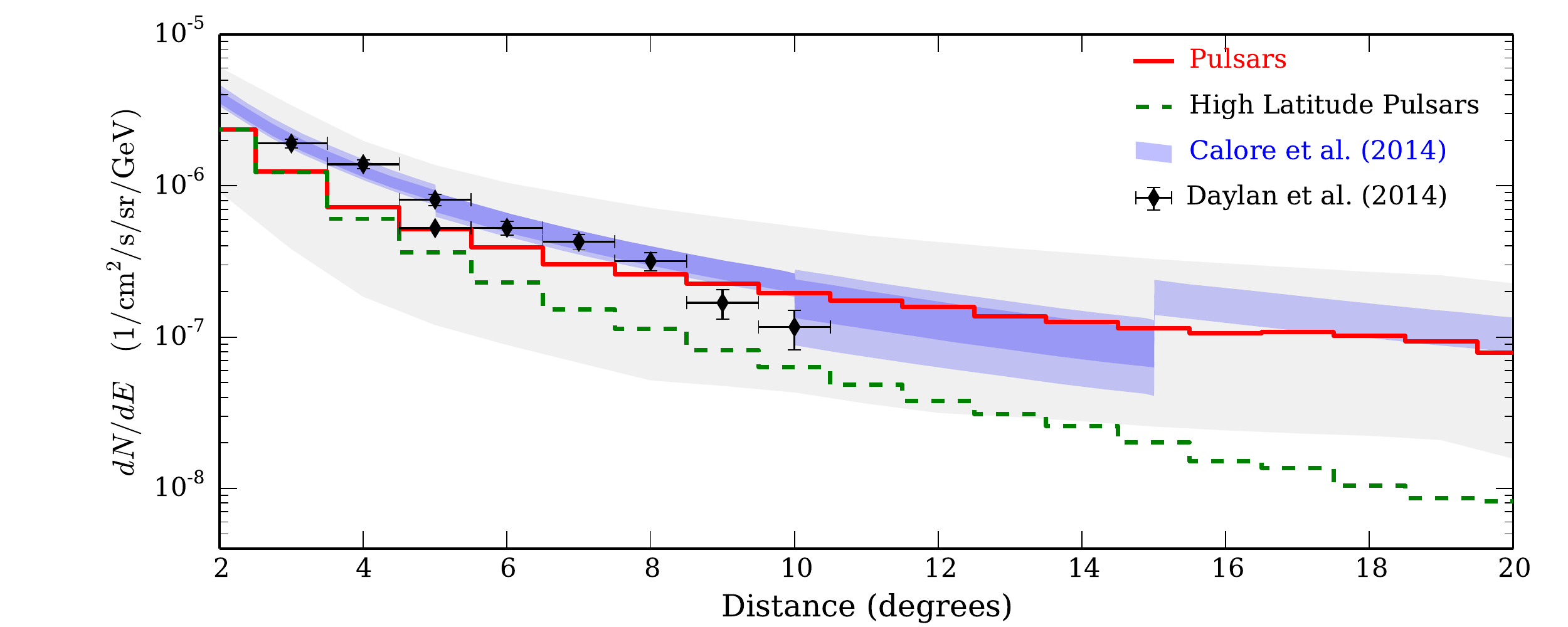}
\caption{\label{fig:daylan} Intensity of 2\,GeV backgrounds
  as a function of angular separation from the Galactic Center. The solid red line shows the average pulsar contribution
  from our fiducial model, obtained from angular rings (excluding the $|b| < 2^\circ$ disk), with the gray shaded region showing the
  systematic uncertainty in the gamma-ray pulsar population.  
The green dashed line is only from pulsars at high Galactic latitude ($|b| > |l|$ and $|b| > 2^\circ$).
Also shown are intensities of the gamma-ray Excess, from Ref.~\cite{2014daylanetal} which used Gaussian angular rings (masking $|b| < 2^\circ$) out to $\approx 10^\circ$ (black points and error bars; the lower point at $5^\circ$ is their ``Full Sky'', similar to their ``Galactic Center'' values) and Ref.~\cite{2014caloreetal} (shaded blue bands) which assume a fixed profile shape.
}
\end{figure*}
%%%%%%%%%%%%%%%%
%%%%%%%%%%%%%%%%

To generate gamma-ray maps of the GC,
we convolve each pulsar with the {\em Fermi} point spread function (PSF)
\citep[see Eq.~(5) of Ref.~][]{2013fermipsf} twice \footnote{This is because the reconstructed images by Refs. \cite{2014daylanetal} and \cite{2014arXiv1409.0042C} have effectively been smoothed twice: once by {\it Fermi}'s PSF and once again when reconstructing the distribution of gamma rays.}. We use luminosity-weighted spectra of the pulsars to estimate their flux density at 2~GeV, as detailed in \cite{2015toappear}.
 In this analysis we remove all pulsars with ages $t \lesssim 10^4\,\yr$, as the gamma-ray
luminosity of the youngest pulsars is poorly understood. To generate the diffuse map, we then subtract sources with fluxes greater than the 50\% completeness limit of {\it Fermi} derived from the brightness distribution of 2FGL point sources between $|b| < 20^\circ$ and $|l|<20^\circ$ \cite{2012ApJS..199...31N}
\begin{equation}
L_{\rm ps} = 1 \times 10^{-10} \frac{1^\circ}{(b^2+(3 ^\circ)^2)^{1/2}}\,{\rm erg\,s}^{-1}\,{\rm cm}^{-2}, \label{eq:ps}
\end{equation}
where $b$ is the Galactic latitude in degrees, as shown in Fig.~\ref{fig:pointsource}. We use 2FGL only to be consistent with \cite{2014arXiv1409.0042C,2014daylanetal}.

In Fig.~\ref{fig:cmz}, we show the surface brightness map of a single
realization of the unresolved gamma-ray pulsars towards the GC.  The flux is
centrally concentrated around the CMZ, but outside of the central few
degrees, pulsars in the Galactic disk become the dominant component,
with a population of point sources extending well above the Galactic
plane.  The outermost regions are the most impacted by Poisson
fluctuations in the underlying pulsar population, and show the
greatest amount of asphericity.  Since most of the flux in the inner $5^\circ$ comes from the CMZ pulsars, flux-weighted measures of the
asymmetry of the emission will not be heavily impacted by the disk.

In Fig.~\ref{fig:daylan}, we compare the surface brightness profile of
young pulsars to the gamma-ray excess at 2~GeV reported by
\citet{2014daylanetal} and
\citet{2014arXiv1409.0042C,2014caloreetal}. The solid red line shows
the mean surface brightness of the pulsars using concentric rings with
$1^\circ$ width that exclude the Galactic plane (at $|b|<2^\circ$), motivated by comparison with \cite{2014daylanetal}, while the
green dashed line includes only the high-latitude pulsars ($|b| > |l|$ and $|b| > 2^\circ$).

We see that the total flux of the young pulsars can plausibly explain much of the GeV Excess in the region we explore in this work, with a slope well within the literature range.  Since pulsar behavior near the gamma-ray death line remains unclear, as is the underlying magnetic field distribution of the pulsars, we have chosen to plot 
along with the total flux implied by Eq.~(\ref{eq:lumfunc}) down to $\dot{E} > 10^{33.5}\,$erg~s$^{-1}$ (solid red line) the range of uncertainty of the underlying model (gray band).  We have done this by varying the main parameters within the empirically observed limits.  We have varied the cutoff between $10^{34} \gtrsim \dot{E} \gtrsim 10^{33}\,$erg~s$^{-1}$, $0.55 \gtrsim \sigma_{\log{B}} \gtrsim 0.30$, and $2.2 \gtrsim C \gtrsim 1.0$.  
Currently, the lowest $\dot{E}$ for a young gamma-ray pulsar is $\sim 3 \times 10^{33}\,$erg s$^{-1}$ \cite{2013pulsarcatalog}. Unfortunately, low-$\dot{E}$ radio-quiet pulsars often lack distances (thus luminosities).  Indeed, tests coupling both local observations and the large-GC angle pulsar population may improve our understanding of these faintest pulsars \cite{2011ApJ...738..114R,2013pulsarcatalog,2014A&A...570A..44H}.

The inferred spectrum of the Excess mostly comes from the $\sim 5^\circ$ region around the GC \cite{2014caloreetal,2014daylanetal}. In our model, the luminosity from this region is dominated by Geminga-like pulsars (which has $\dot{E} \approx 10^{34.5}\,$erg s$^{-1}$).  Fig.~\ref{fig:geminga} shows the flux-weighted spectrum of all pulsars that are beyond the plane ($|b| > 2^\circ$) and within $5^\circ$ of the GC, which is comparable to the Excess spectral shape near the peak.  Note that $\gtrsim 10\,$GeV pulsar emission may be underestimated by using an exponential cut-off \cite[e.g.,][]{2013pulsarcatalog,2014arXiv1407.5583C,2015toappear}.

\textbf{Discussion.}---
We have shown that a population of 
young ordinary pulsars originating in the central molecular zone of the Galactic center and the Galactic disk naturally produces a gamma-ray spectrum, amplitude, and
spatial distribution resembling the observed GeV excess in the Galactic center. 
Our model uses the observed properties of radio
\cite{2005MNRAS.360..974H,fauchergiguerekaspi2006} and gamma-ray
pulsars \cite{2013pulsarcatalog} to estimate their gamma-ray flux. 
The young pulsar luminosity profile is then determined by kicked pulsars evolving with age, while
we include no contribution from millisecond pulsars.

 \citet{2014caloreetal} found that the Excess brightness varied
 between different analyses at the 3$-\sigma$ level, suggesting at least a
 factor of two uncertainty in its absolute normalization.  In this
 work, we used a common, simplified model for pulsar evolution. This
 model also has a factor of two uncertainty due to discrepancies in the
 preferred surface magnetic fields for radio pulsars and gamma-ray
 pulsars,  pulsar behavior near the
 gamma-ray death line, and the GC SN rate (though our fiducial value is near that
 preferred in Fermi Bubble models \cite{2011MNRAS.413..763C,2014MNRAS.444L..39L}).  
We discuss these in \cite{2015toappear}, although we note here that 
 the parameters used by our models were specifically optimized to
 reproduce the currently observed distribution of pulsars.

Nevertheless, there remains much room for improvement.  While current ``gap'' models of pulsar emission \cite[e.g.,][]{1975ApJ...196...51R,1983ApJ...266..215A,1986ApJ...300..500C,1996ApJ...470..469R,2008ApJ...680.1378H} work reasonably well \cite{2015A&A...575A...3P}, ab initio models under development promise better physical understanding \cite{2010ApJ...715.1282B,2014ApJ...793...97K,2014ApJ...795L..22C,2014arXiv1412.0673P}.  Ideally, we would also have a similar understanding of neutron star birth kicks, spins, masses, and magnetic fields \cite{2012MNRAS.423.1805N,2013ApJ...768..115O,2013A&A...552A.126W,2014arXiv1409.5779B}.

We discuss further observations to help determine the young pulsar fraction in \cite{2015toappear}.
These include the pulsar contribution to a similar GeV excess seen along the Galactic disk \cite{2012ApJ...750....3A}, which may also be evident as residual emission in the inner Galaxy \cite{MurgiaTalk}.
We have not here included electrons and positrons from pulsar winds (e.g., Geminga displays diffuse TeV emission \cite{2009ApJ...700L.127A}  due to $e^\pm$ \cite{yuksel2009}).  If proportional to $\dot{E}$, most of the output is concentrated near the plane to yield an inverse Compton flux (as may be needed in the inner galaxy \cite{2014arXiv1409.0042C,MurgiaTalk}).

Despite present uncertainties, young pulsars must be accounted for in
future models of the Excess, with our conservative estimates suggesting that they constitute a substantial fraction.   If the Excess extends to higher energies, as in \cite{2014caloreetal}, and if pulsars only contribute a fraction at GeV energies, a higher-mass dark matter candidate may yet be present (see \cite{2015toappear}). 
Unlike the GC, old dwarf galaxies should not contain young pulsars, so improved 
gamma-ray studies \cite{2015arXiv150302641F,2014arXiv1410.2242G}  will help our understanding of both the Excess and pulsar physics.
 Clearly, additional {\it Fermi} data will be vital to resolve more sources, characterize pulsars, and accumulate statistics at the highest energies, where upcoming IACTs will also be relevant \cite{2013APh....43..189D,2013arXiv1305.0302W,2013arXiv1305.0082W}, to dissect the Excess.

%\vspace{-0.35in}
%\acknowledgments
%\vspace{-0.2in}
\textit{Acknowledgments:}
We thank John Bally, Casey Law, and Jon Mauerhan for useful discussions on GC star formation and SN rates; Seth Digel, Troy Porter, Roger Romani, and Matt Wood on {\it Fermi} and pulsars.
RMO acknowledges the support provided by NSF grant AST-1313021 and the hospitality of the Aspen Center for Physics.
MDK acknowledges support provided by Department of Energy contract DE-AC02-76SF00515, and the KIPAC Kavli Fellowship made possible by The Kavli Foundation.

\bibliography{pulsars}

\begin{thebibliography}{105}
\expandafter\ifx\csname natexlab\endcsname\relax\def\natexlab#1{#1}\fi
\expandafter\ifx\csname bibnamefont\endcsname\relax
  \def\bibnamefont#1{#1}\fi
\expandafter\ifx\csname bibfnamefont\endcsname\relax
  \def\bibfnamefont#1{#1}\fi
\expandafter\ifx\csname citenamefont\endcsname\relax
  \def\citenamefont#1{#1}\fi
\expandafter\ifx\csname url\endcsname\relax
  \def\url#1{\texttt{#1}}\fi
\expandafter\ifx\csname urlprefix\endcsname\relax\def\urlprefix{URL }\fi
\providecommand{\bibinfo}[2]{#2}
\providecommand{\eprint}[2][]{\url{#2}}

\bibitem[{\citenamefont{{Atwood} et~al.}(2009)\citenamefont{{Atwood}, {Abdo},
  {Ackermann}, {Althouse}, {Anderson}, {Axelsson}, {Baldini}, {Ballet}, {Band},
  {Barbiellini} et~al.}}]{2009ApJ...697.1071A}
\bibinfo{author}{\bibfnamefont{W.~B.} \bibnamefont{{Atwood}}},
  \bibinfo{author}{\bibfnamefont{A.~A.} \bibnamefont{{Abdo}}},
  \bibinfo{author}{\bibfnamefont{M.}~\bibnamefont{{Ackermann}}},
  \bibinfo{author}{\bibfnamefont{W.}~\bibnamefont{{Althouse}}},
  \bibinfo{author}{\bibfnamefont{B.}~\bibnamefont{{Anderson}}},
  \bibinfo{author}{\bibfnamefont{M.}~\bibnamefont{{Axelsson}}},
  \bibinfo{author}{\bibfnamefont{L.}~\bibnamefont{{Baldini}}},
  \bibinfo{author}{\bibfnamefont{J.}~\bibnamefont{{Ballet}}},
  \bibinfo{author}{\bibfnamefont{D.~L.} \bibnamefont{{Band}}},
  \bibinfo{author}{\bibfnamefont{G.}~\bibnamefont{{Barbiellini}}},
  \bibnamefont{et~al.}, \bibinfo{journal}{\apj} \textbf{\bibinfo{volume}{697}},
  \bibinfo{pages}{1071} (\bibinfo{year}{2009}), \eprint{0902.1089}.

\bibitem[{\citenamefont{{Caraveo}}(2014)}]{2014ARA&A..52..211C}
\bibinfo{author}{\bibfnamefont{P.~A.} \bibnamefont{{Caraveo}}},
  \bibinfo{journal}{\araa} \textbf{\bibinfo{volume}{52}}, \bibinfo{pages}{211}
  (\bibinfo{year}{2014}), \eprint{1312.2913}.

\bibitem[{\citenamefont{{Abdo} et~al.}(2013)\citenamefont{{Abdo}, {Ajello},
  {Allafort}, {Baldini}, {Ballet}, {Barbiellini}, {Baring}, {Bastieri},
  {Belfiore}, {Bellazzini} et~al.}}]{2013pulsarcatalog}
\bibinfo{author}{\bibfnamefont{A.~A.} \bibnamefont{{Abdo}}},
  \bibinfo{author}{\bibfnamefont{M.}~\bibnamefont{{Ajello}}},
  \bibinfo{author}{\bibfnamefont{A.}~\bibnamefont{{Allafort}}},
  \bibinfo{author}{\bibfnamefont{L.}~\bibnamefont{{Baldini}}},
  \bibinfo{author}{\bibfnamefont{J.}~\bibnamefont{{Ballet}}},
  \bibinfo{author}{\bibfnamefont{G.}~\bibnamefont{{Barbiellini}}},
  \bibinfo{author}{\bibfnamefont{M.~G.} \bibnamefont{{Baring}}},
  \bibinfo{author}{\bibfnamefont{D.}~\bibnamefont{{Bastieri}}},
  \bibinfo{author}{\bibfnamefont{A.}~\bibnamefont{{Belfiore}}},
  \bibinfo{author}{\bibfnamefont{R.}~\bibnamefont{{Bellazzini}}},
  \bibnamefont{et~al.}, \bibinfo{journal}{\apjs}
  \textbf{\bibinfo{volume}{208}}, \bibinfo{eid}{17} (\bibinfo{year}{2013}),
  \eprint{1305.4385}.

\bibitem[{\citenamefont{{Laffon} et~al.}(2015)\citenamefont{{Laffon}, {Smith},
  {Guillemot}, and {for the Fermi-LAT Collaboration}}}]{2015arXiv150203251L}
\bibinfo{author}{\bibfnamefont{H.}~\bibnamefont{{Laffon}}},
  \bibinfo{author}{\bibfnamefont{D.~A.} \bibnamefont{{Smith}}},
  \bibinfo{author}{\bibfnamefont{L.}~\bibnamefont{{Guillemot}}},
  \bibnamefont{and} \bibinfo{author}{\bibnamefont{{for the Fermi-LAT
  Collaboration}}}, \bibinfo{journal}{ArXiv e-prints}  (\bibinfo{year}{2015}),
  \eprint{1502.03251}.

\bibitem[{\citenamefont{{Watters} and {Romani}}(2011)}]{2011ApJ...727..123W}
\bibinfo{author}{\bibfnamefont{K.~P.} \bibnamefont{{Watters}}}
  \bibnamefont{and} \bibinfo{author}{\bibfnamefont{R.~W.}
  \bibnamefont{{Romani}}}, \bibinfo{journal}{\apj}
  \textbf{\bibinfo{volume}{727}}, \bibinfo{eid}{123} (\bibinfo{year}{2011}),
  \eprint{1009.5305}.

\bibitem[{\citenamefont{{Pierbattista}
  et~al.}(2012)\citenamefont{{Pierbattista}, {Grenier}, {Harding}, and
  {Gonthier}}}]{2012A&A...545A..42P}
\bibinfo{author}{\bibfnamefont{M.}~\bibnamefont{{Pierbattista}}},
  \bibinfo{author}{\bibfnamefont{I.~A.} \bibnamefont{{Grenier}}},
  \bibinfo{author}{\bibfnamefont{A.~K.} \bibnamefont{{Harding}}},
  \bibnamefont{and} \bibinfo{author}{\bibfnamefont{P.~L.}
  \bibnamefont{{Gonthier}}}, \bibinfo{journal}{\aap}
  \textbf{\bibinfo{volume}{545}}, \bibinfo{eid}{A42} (\bibinfo{year}{2012}),
  \eprint{1206.5634}.

\bibitem[{\citenamefont{{Halpern} and {Holt}}(1992)}]{1992Natur.357..222H}
\bibinfo{author}{\bibfnamefont{J.~P.} \bibnamefont{{Halpern}}}
  \bibnamefont{and} \bibinfo{author}{\bibfnamefont{S.~S.}
  \bibnamefont{{Holt}}}, \bibinfo{journal}{\nat}
  \textbf{\bibinfo{volume}{357}}, \bibinfo{pages}{222} (\bibinfo{year}{1992}).

\bibitem[{\citenamefont{{Bertsch} et~al.}(1992)\citenamefont{{Bertsch},
  {Brazier}, {Fichtel}, {Hartman}, {Hunter}, {Kanbach}, {Kniffen}, {Kwok},
  {Lin}, and {Mattox}}}]{1992Natur.357..306B}
\bibinfo{author}{\bibfnamefont{D.~L.} \bibnamefont{{Bertsch}}},
  \bibinfo{author}{\bibfnamefont{K.~T.~S.} \bibnamefont{{Brazier}}},
  \bibinfo{author}{\bibfnamefont{C.~E.} \bibnamefont{{Fichtel}}},
  \bibinfo{author}{\bibfnamefont{R.~C.} \bibnamefont{{Hartman}}},
  \bibinfo{author}{\bibfnamefont{S.~D.} \bibnamefont{{Hunter}}},
  \bibinfo{author}{\bibfnamefont{G.}~\bibnamefont{{Kanbach}}},
  \bibinfo{author}{\bibfnamefont{D.~A.} \bibnamefont{{Kniffen}}},
  \bibinfo{author}{\bibfnamefont{P.~W.} \bibnamefont{{Kwok}}},
  \bibinfo{author}{\bibfnamefont{Y.~C.} \bibnamefont{{Lin}}}, \bibnamefont{and}
  \bibinfo{author}{\bibfnamefont{J.~R.} \bibnamefont{{Mattox}}},
  \bibinfo{journal}{\nat} \textbf{\bibinfo{volume}{357}}, \bibinfo{pages}{306}
  (\bibinfo{year}{1992}).

\bibitem[{\citenamefont{{Jungman} et~al.}(1996)\citenamefont{{Jungman},
  {Kamionkowski}, and {Griest}}}]{1996PhR...267..195J}
\bibinfo{author}{\bibfnamefont{G.}~\bibnamefont{{Jungman}}},
  \bibinfo{author}{\bibfnamefont{M.}~\bibnamefont{{Kamionkowski}}},
  \bibnamefont{and} \bibinfo{author}{\bibfnamefont{K.}~\bibnamefont{{Griest}}},
  \bibinfo{journal}{\physrep} \textbf{\bibinfo{volume}{267}},
  \bibinfo{pages}{195} (\bibinfo{year}{1996}), \eprint{hep-ph/9506380}.

\bibitem[{\citenamefont{{Gondolo} et~al.}(2004)\citenamefont{{Gondolo},
  {Edsj{\"o}}, {Ullio}, {Bergstr{\"o}m}, {Schelke}, and
  {Baltz}}}]{2004JCAP...07..008G}
\bibinfo{author}{\bibfnamefont{P.}~\bibnamefont{{Gondolo}}},
  \bibinfo{author}{\bibfnamefont{J.}~\bibnamefont{{Edsj{\"o}}}},
  \bibinfo{author}{\bibfnamefont{P.}~\bibnamefont{{Ullio}}},
  \bibinfo{author}{\bibfnamefont{L.}~\bibnamefont{{Bergstr{\"o}m}}},
  \bibinfo{author}{\bibfnamefont{M.}~\bibnamefont{{Schelke}}},
  \bibnamefont{and} \bibinfo{author}{\bibfnamefont{E.~A.}
  \bibnamefont{{Baltz}}}, \bibinfo{journal}{\jcap}
  \textbf{\bibinfo{volume}{7}}, \bibinfo{eid}{008} (\bibinfo{year}{2004}),
  \eprint{astro-ph/0406204}.

\bibitem[{\citenamefont{{Bertone} et~al.}(2005)\citenamefont{{Bertone},
  {Hooper}, and {Silk}}}]{2005PhR...405..279B}
\bibinfo{author}{\bibfnamefont{G.}~\bibnamefont{{Bertone}}},
  \bibinfo{author}{\bibfnamefont{D.}~\bibnamefont{{Hooper}}}, \bibnamefont{and}
  \bibinfo{author}{\bibfnamefont{J.}~\bibnamefont{{Silk}}},
  \bibinfo{journal}{\physrep} \textbf{\bibinfo{volume}{405}},
  \bibinfo{pages}{279} (\bibinfo{year}{2005}), \eprint{hep-ph/0404175}.

\bibitem[{\citenamefont{{Strigari}}(2013)}]{2013PhR...531....1S}
\bibinfo{author}{\bibfnamefont{L.~E.} \bibnamefont{{Strigari}}},
  \bibinfo{journal}{\physrep} \textbf{\bibinfo{volume}{531}},
  \bibinfo{pages}{1} (\bibinfo{year}{2013}), \eprint{1211.7090}.

\bibitem[{\citenamefont{{Ackermann} et~al.}(2014)\citenamefont{{Ackermann},
  {Albert}, {Anderson}, {Baldini}, {Ballet}, {Barbiellini}, {Bastieri},
  {Bechtol}, {Bellazzini}, {Bissaldi} et~al.}}]{2014PhRvD..89d2001A}
\bibinfo{author}{\bibfnamefont{M.}~\bibnamefont{{Ackermann}}},
  \bibinfo{author}{\bibfnamefont{A.}~\bibnamefont{{Albert}}},
  \bibinfo{author}{\bibfnamefont{B.}~\bibnamefont{{Anderson}}},
  \bibinfo{author}{\bibfnamefont{L.}~\bibnamefont{{Baldini}}},
  \bibinfo{author}{\bibfnamefont{J.}~\bibnamefont{{Ballet}}},
  \bibinfo{author}{\bibfnamefont{G.}~\bibnamefont{{Barbiellini}}},
  \bibinfo{author}{\bibfnamefont{D.}~\bibnamefont{{Bastieri}}},
  \bibinfo{author}{\bibfnamefont{K.}~\bibnamefont{{Bechtol}}},
  \bibinfo{author}{\bibfnamefont{R.}~\bibnamefont{{Bellazzini}}},
  \bibinfo{author}{\bibfnamefont{E.}~\bibnamefont{{Bissaldi}}},
  \bibnamefont{et~al.}, \bibinfo{journal}{\prd} \textbf{\bibinfo{volume}{89}},
  \bibinfo{eid}{042001} (\bibinfo{year}{2014}).

\bibitem[{\citenamefont{{Fermi-LAT Collaboration}}(2015)}]{2015arXiv150302641F}
\bibinfo{author}{\bibnamefont{{Fermi-LAT Collaboration}}},
  \bibinfo{journal}{ArXiv e-prints}  (\bibinfo{year}{2015}),
  \eprint{1503.02641}.

\bibitem[{\citenamefont{{Zwicky}}(1933)}]{1933AcHPh...6..110Z}
\bibinfo{author}{\bibfnamefont{F.}~\bibnamefont{{Zwicky}}},
  \bibinfo{journal}{Helvetica Physica Acta} \textbf{\bibinfo{volume}{6}},
  \bibinfo{pages}{110} (\bibinfo{year}{1933}).

\bibitem[{\citenamefont{{Planck Collaboration}}(2015)}]{2015arXiv150201589P}
\bibinfo{author}{\bibnamefont{{Planck Collaboration}}}, \bibinfo{journal}{ArXiv
  e-prints}  (\bibinfo{year}{2015}), \eprint{1502.01589}.

\bibitem[{\citenamefont{{Goodenough} and {Hooper}}(2009)}]{2009arXiv0910.2998G}
\bibinfo{author}{\bibfnamefont{L.}~\bibnamefont{{Goodenough}}}
  \bibnamefont{and} \bibinfo{author}{\bibfnamefont{D.}~\bibnamefont{{Hooper}}},
  \bibinfo{journal}{ArXiv e-prints}  (\bibinfo{year}{2009}),
  \eprint{0910.2998}.

\bibitem[{\citenamefont{{Hooper} and {Goodenough}}(2011)}]{2011PhLB..697..412H}
\bibinfo{author}{\bibfnamefont{D.}~\bibnamefont{{Hooper}}} \bibnamefont{and}
  \bibinfo{author}{\bibfnamefont{L.}~\bibnamefont{{Goodenough}}},
  \bibinfo{journal}{Physics Letters B} \textbf{\bibinfo{volume}{697}},
  \bibinfo{pages}{412} (\bibinfo{year}{2011}), \eprint{1010.2752}.

\bibitem[{\citenamefont{{Hooper} and {Linden}}(2011)}]{2011PhRvD..84l3005H}
\bibinfo{author}{\bibfnamefont{D.}~\bibnamefont{{Hooper}}} \bibnamefont{and}
  \bibinfo{author}{\bibfnamefont{T.}~\bibnamefont{{Linden}}},
  \bibinfo{journal}{\prd} \textbf{\bibinfo{volume}{84}}, \bibinfo{eid}{123005}
  (\bibinfo{year}{2011}), \eprint{1110.0006}.

\bibitem[{\citenamefont{{Abazajian} and
  {Kaplinghat}}(2012)}]{2012PhRvD..86h3511A}
\bibinfo{author}{\bibfnamefont{K.~N.} \bibnamefont{{Abazajian}}}
  \bibnamefont{and}
  \bibinfo{author}{\bibfnamefont{M.}~\bibnamefont{{Kaplinghat}}},
  \bibinfo{journal}{\prd} \textbf{\bibinfo{volume}{86}}, \bibinfo{eid}{083511}
  (\bibinfo{year}{2012}), \eprint{1207.6047}.

\bibitem[{\citenamefont{{Hooper} and {Slatyer}}(2013)}]{2013PDU.....2..118H}
\bibinfo{author}{\bibfnamefont{D.}~\bibnamefont{{Hooper}}} \bibnamefont{and}
  \bibinfo{author}{\bibfnamefont{T.~R.} \bibnamefont{{Slatyer}}},
  \bibinfo{journal}{Physics of the Dark Universe} \textbf{\bibinfo{volume}{2}},
  \bibinfo{pages}{118} (\bibinfo{year}{2013}), \eprint{1302.6589}.

\bibitem[{\citenamefont{{Gordon} and
  {Mac{\'{\i}}as}}(2013)}]{2013PhRvD..88h3521G}
\bibinfo{author}{\bibfnamefont{C.}~\bibnamefont{{Gordon}}} \bibnamefont{and}
  \bibinfo{author}{\bibfnamefont{O.}~\bibnamefont{{Mac{\'{\i}}as}}},
  \bibinfo{journal}{\prd} \textbf{\bibinfo{volume}{88}}, \bibinfo{eid}{083521}
  (\bibinfo{year}{2013}), \eprint{1306.5725}.

\bibitem[{\citenamefont{{Macias} and {Gordon}}(2014)}]{2014PhRvD..89f3515M}
\bibinfo{author}{\bibfnamefont{O.}~\bibnamefont{{Macias}}} \bibnamefont{and}
  \bibinfo{author}{\bibfnamefont{C.}~\bibnamefont{{Gordon}}},
  \bibinfo{journal}{\prd} \textbf{\bibinfo{volume}{89}}, \bibinfo{eid}{063515}
  (\bibinfo{year}{2014}), \eprint{1312.6671}.

\bibitem[{\citenamefont{{Abazajian} et~al.}(2014)\citenamefont{{Abazajian},
  {Canac}, {Horiuchi}, and {Kaplinghat}}}]{2014PhRvD..90b3526A}
\bibinfo{author}{\bibfnamefont{K.~N.} \bibnamefont{{Abazajian}}},
  \bibinfo{author}{\bibfnamefont{N.}~\bibnamefont{{Canac}}},
  \bibinfo{author}{\bibfnamefont{S.}~\bibnamefont{{Horiuchi}}},
  \bibnamefont{and}
  \bibinfo{author}{\bibfnamefont{M.}~\bibnamefont{{Kaplinghat}}},
  \bibinfo{journal}{\prd} \textbf{\bibinfo{volume}{90}}, \bibinfo{eid}{023526}
  (\bibinfo{year}{2014}), \eprint{1402.4090}.

\bibitem[{\citenamefont{{Daylan} et~al.}(2014)\citenamefont{{Daylan},
  {Finkbeiner}, {Hooper}, {Linden}, {Portillo}, {Rodd}, and
  {Slatyer}}}]{2014daylanetal}
\bibinfo{author}{\bibfnamefont{T.}~\bibnamefont{{Daylan}}},
  \bibinfo{author}{\bibfnamefont{D.~P.} \bibnamefont{{Finkbeiner}}},
  \bibinfo{author}{\bibfnamefont{D.}~\bibnamefont{{Hooper}}},
  \bibinfo{author}{\bibfnamefont{T.}~\bibnamefont{{Linden}}},
  \bibinfo{author}{\bibfnamefont{S.~K.~N.} \bibnamefont{{Portillo}}},
  \bibinfo{author}{\bibfnamefont{N.~L.} \bibnamefont{{Rodd}}},
  \bibnamefont{and} \bibinfo{author}{\bibfnamefont{T.~R.}
  \bibnamefont{{Slatyer}}}, \bibinfo{journal}{ArXiv e-prints}
  (\bibinfo{year}{2014}), \eprint{1402.6703}.

\bibitem[{\citenamefont{{Zhou} et~al.}(2014)\citenamefont{{Zhou}, {Liang},
  {Huang}, {Li}, {Fan}, {Feng}, and {Chang}}}]{2014arXiv1406.6948Z}
\bibinfo{author}{\bibfnamefont{B.}~\bibnamefont{{Zhou}}},
  \bibinfo{author}{\bibfnamefont{Y.-F.} \bibnamefont{{Liang}}},
  \bibinfo{author}{\bibfnamefont{X.}~\bibnamefont{{Huang}}},
  \bibinfo{author}{\bibfnamefont{X.}~\bibnamefont{{Li}}},
  \bibinfo{author}{\bibfnamefont{Y.-Z.} \bibnamefont{{Fan}}},
  \bibinfo{author}{\bibfnamefont{L.}~\bibnamefont{{Feng}}}, \bibnamefont{and}
  \bibinfo{author}{\bibfnamefont{J.}~\bibnamefont{{Chang}}},
  \bibinfo{journal}{ArXiv e-prints}  (\bibinfo{year}{2014}),
  \eprint{1406.6948}.

\bibitem[{\citenamefont{{Calore}
  et~al.}(2015{\natexlab{a}})\citenamefont{{Calore}, {Cholis}, and
  {Weniger}}}]{2014arXiv1409.0042C}
\bibinfo{author}{\bibfnamefont{F.}~\bibnamefont{{Calore}}},
  \bibinfo{author}{\bibfnamefont{I.}~\bibnamefont{{Cholis}}}, \bibnamefont{and}
  \bibinfo{author}{\bibfnamefont{C.}~\bibnamefont{{Weniger}}},
  \bibinfo{journal}{\jcap} \textbf{\bibinfo{volume}{3}}, \bibinfo{eid}{038}
  (\bibinfo{year}{2015}{\natexlab{a}}), \eprint{1409.0042}.

\bibitem[{\citenamefont{{Calore}
  et~al.}(2015{\natexlab{b}})\citenamefont{{Calore}, {Cholis}, {McCabe}, and
  {Weniger}}}]{2014caloreetal}
\bibinfo{author}{\bibfnamefont{F.}~\bibnamefont{{Calore}}},
  \bibinfo{author}{\bibfnamefont{I.}~\bibnamefont{{Cholis}}},
  \bibinfo{author}{\bibfnamefont{C.}~\bibnamefont{{McCabe}}}, \bibnamefont{and}
  \bibinfo{author}{\bibfnamefont{C.}~\bibnamefont{{Weniger}}},
  \bibinfo{journal}{\prd} \textbf{\bibinfo{volume}{91}}, \bibinfo{eid}{063003}
  (\bibinfo{year}{2015}{\natexlab{b}}), \eprint{1411.4647}.

\bibitem[{\citenamefont{{Murgia}}(2014)}]{MurgiaTalk}
\bibinfo{author}{\bibfnamefont{S.}~\bibnamefont{{Murgia}}}
  (\bibinfo{year}{2014}), \bibinfo{note}{"{\it Observation of the high energy
  gamma-ray emission towards the Galactic center}", Talk at Fifth Fermi
  Symposium, Nagoya, October 2014}.

\bibitem[{\citenamefont{{Abdullah} et~al.}(2014)\citenamefont{{Abdullah},
  {DiFranzo}, {Rajaraman}, {Tait}, {Tanedo}, and
  {Wijangco}}}]{2014PhRvD..90c5004A}
\bibinfo{author}{\bibfnamefont{M.}~\bibnamefont{{Abdullah}}},
  \bibinfo{author}{\bibfnamefont{A.}~\bibnamefont{{DiFranzo}}},
  \bibinfo{author}{\bibfnamefont{A.}~\bibnamefont{{Rajaraman}}},
  \bibinfo{author}{\bibfnamefont{T.~M.~P.} \bibnamefont{{Tait}}},
  \bibinfo{author}{\bibfnamefont{P.}~\bibnamefont{{Tanedo}}}, \bibnamefont{and}
  \bibinfo{author}{\bibfnamefont{A.~M.} \bibnamefont{{Wijangco}}},
  \bibinfo{journal}{\prd} \textbf{\bibinfo{volume}{90}}, \bibinfo{eid}{035004}
  (\bibinfo{year}{2014}), \eprint{1404.6528}.

\bibitem[{\citenamefont{{Izaguirre} et~al.}(2014)\citenamefont{{Izaguirre},
  {Krnjaic}, and {Shuve}}}]{2014PhRvD..90e5002I}
\bibinfo{author}{\bibfnamefont{E.}~\bibnamefont{{Izaguirre}}},
  \bibinfo{author}{\bibfnamefont{G.}~\bibnamefont{{Krnjaic}}},
  \bibnamefont{and} \bibinfo{author}{\bibfnamefont{B.}~\bibnamefont{{Shuve}}},
  \bibinfo{journal}{\prd} \textbf{\bibinfo{volume}{90}}, \bibinfo{eid}{055002}
  (\bibinfo{year}{2014}), \eprint{1404.2018}.

\bibitem[{\citenamefont{{Ipek} et~al.}(2014)\citenamefont{{Ipek}, {McKeen}, and
  {Nelson}}}]{2014PhRvD..90e5021I}
\bibinfo{author}{\bibfnamefont{S.}~\bibnamefont{{Ipek}}},
  \bibinfo{author}{\bibfnamefont{D.}~\bibnamefont{{McKeen}}}, \bibnamefont{and}
  \bibinfo{author}{\bibfnamefont{A.~E.} \bibnamefont{{Nelson}}},
  \bibinfo{journal}{\prd} \textbf{\bibinfo{volume}{90}}, \bibinfo{eid}{055021}
  (\bibinfo{year}{2014}), \eprint{1404.3716}.

\bibitem[{\citenamefont{{Kong} and {Park}}(2014)}]{2014NuPhB.888..154K}
\bibinfo{author}{\bibfnamefont{K.}~\bibnamefont{{Kong}}} \bibnamefont{and}
  \bibinfo{author}{\bibfnamefont{J.-C.} \bibnamefont{{Park}}},
  \bibinfo{journal}{Nuclear Physics B} \textbf{\bibinfo{volume}{888}},
  \bibinfo{pages}{154} (\bibinfo{year}{2014}), \eprint{1404.3741}.

\bibitem[{\citenamefont{{Ko} et~al.}(2014)\citenamefont{{Ko}, {Park}, and
  {Tang}}}]{2014JCAP...09..013K}
\bibinfo{author}{\bibfnamefont{P.}~\bibnamefont{{Ko}}},
  \bibinfo{author}{\bibfnamefont{W.-I.} \bibnamefont{{Park}}},
  \bibnamefont{and} \bibinfo{author}{\bibfnamefont{Y.}~\bibnamefont{{Tang}}},
  \bibinfo{journal}{\jcap} \textbf{\bibinfo{volume}{9}}, \bibinfo{eid}{013}
  (\bibinfo{year}{2014}), \eprint{1404.5257}.

\bibitem[{\citenamefont{{Boehm} et~al.}(2014)\citenamefont{{Boehm}, {Dolan},
  and {McCabe}}}]{2014PhRvD..90b3531B}
\bibinfo{author}{\bibfnamefont{C.}~\bibnamefont{{Boehm}}},
  \bibinfo{author}{\bibfnamefont{M.~J.} \bibnamefont{{Dolan}}},
  \bibnamefont{and} \bibinfo{author}{\bibfnamefont{C.}~\bibnamefont{{McCabe}}},
  \bibinfo{journal}{\prd} \textbf{\bibinfo{volume}{90}}, \bibinfo{eid}{023531}
  (\bibinfo{year}{2014}).

\bibitem[{\citenamefont{{Cheung} et~al.}(2014)\citenamefont{{Cheung},
  {Papucci}, {Sanford}, {Shah}, and {Zurek}}}]{2014PhRvD..90g5011C}
\bibinfo{author}{\bibfnamefont{C.}~\bibnamefont{{Cheung}}},
  \bibinfo{author}{\bibfnamefont{M.}~\bibnamefont{{Papucci}}},
  \bibinfo{author}{\bibfnamefont{D.}~\bibnamefont{{Sanford}}},
  \bibinfo{author}{\bibfnamefont{N.~R.} \bibnamefont{{Shah}}},
  \bibnamefont{and} \bibinfo{author}{\bibfnamefont{K.~M.}
  \bibnamefont{{Zurek}}}, \bibinfo{journal}{\prd}
  \textbf{\bibinfo{volume}{90}}, \bibinfo{eid}{075011} (\bibinfo{year}{2014}),
  \eprint{1406.6372}.

\bibitem[{\citenamefont{{Freytsis} et~al.}(2014)\citenamefont{{Freytsis},
  {Robinson}, and {Tsai}}}]{2014arXiv1410.3818F}
\bibinfo{author}{\bibfnamefont{M.}~\bibnamefont{{Freytsis}}},
  \bibinfo{author}{\bibfnamefont{D.~J.} \bibnamefont{{Robinson}}},
  \bibnamefont{and} \bibinfo{author}{\bibfnamefont{Y.}~\bibnamefont{{Tsai}}},
  \bibinfo{journal}{ArXiv e-prints}  (\bibinfo{year}{2014}),
  \eprint{1410.3818}.

\bibitem[{\citenamefont{{Agrawal} et~al.}(2014)\citenamefont{{Agrawal},
  {Batell}, {Fox}, and {Harnik}}}]{2014arXiv1411.2592A}
\bibinfo{author}{\bibfnamefont{P.}~\bibnamefont{{Agrawal}}},
  \bibinfo{author}{\bibfnamefont{B.}~\bibnamefont{{Batell}}},
  \bibinfo{author}{\bibfnamefont{P.~J.} \bibnamefont{{Fox}}}, \bibnamefont{and}
  \bibinfo{author}{\bibfnamefont{R.}~\bibnamefont{{Harnik}}},
  \bibinfo{journal}{ArXiv e-prints}  (\bibinfo{year}{2014}),
  \eprint{1411.2592}.

\bibitem[{\citenamefont{{Martin} et~al.}(2014)\citenamefont{{Martin},
  {Shelton}, and {Unwin}}}]{2014PhRvD..90j3513M}
\bibinfo{author}{\bibfnamefont{A.}~\bibnamefont{{Martin}}},
  \bibinfo{author}{\bibfnamefont{J.}~\bibnamefont{{Shelton}}},
  \bibnamefont{and} \bibinfo{author}{\bibfnamefont{J.}~\bibnamefont{{Unwin}}},
  \bibinfo{journal}{\prd} \textbf{\bibinfo{volume}{90}}, \bibinfo{eid}{103513}
  (\bibinfo{year}{2014}), \eprint{1405.0272}.

\bibitem[{\citenamefont{{Liu} et~al.}(2014)\citenamefont{{Liu}, {Weiner}, and
  {Xue}}}]{2014arXiv1412.1485L}
\bibinfo{author}{\bibfnamefont{J.}~\bibnamefont{{Liu}}},
  \bibinfo{author}{\bibfnamefont{N.}~\bibnamefont{{Weiner}}}, \bibnamefont{and}
  \bibinfo{author}{\bibfnamefont{W.}~\bibnamefont{{Xue}}},
  \bibinfo{journal}{ArXiv e-prints}  (\bibinfo{year}{2014}),
  \eprint{1412.1485}.

\bibitem[{\citenamefont{{Bell} et~al.}(2015)\citenamefont{{Bell}, {Horiuchi},
  and {Shoemaker}}}]{2015PhRvD..91b3505B}
\bibinfo{author}{\bibfnamefont{N.~F.} \bibnamefont{{Bell}}},
  \bibinfo{author}{\bibfnamefont{S.}~\bibnamefont{{Horiuchi}}},
  \bibnamefont{and} \bibinfo{author}{\bibfnamefont{I.~M.}
  \bibnamefont{{Shoemaker}}}, \bibinfo{journal}{\prd}
  \textbf{\bibinfo{volume}{91}}, \bibinfo{eid}{023505} (\bibinfo{year}{2015}),
  \eprint{1408.5142}.

\bibitem[{\citenamefont{{Cahill-Rowley}
  et~al.}(2015)\citenamefont{{Cahill-Rowley}, {Gainer}, {Hewett}, and
  {Rizzo}}}]{2015JHEP...02..057C}
\bibinfo{author}{\bibfnamefont{M.}~\bibnamefont{{Cahill-Rowley}}},
  \bibinfo{author}{\bibfnamefont{J.~S.} \bibnamefont{{Gainer}}},
  \bibinfo{author}{\bibfnamefont{J.~L.} \bibnamefont{{Hewett}}},
  \bibnamefont{and} \bibinfo{author}{\bibfnamefont{T.~G.}
  \bibnamefont{{Rizzo}}}, \bibinfo{journal}{Journal of High Energy Physics}
  \textbf{\bibinfo{volume}{2}}, \bibinfo{pages}{57} (\bibinfo{year}{2015}),
  \eprint{1409.1573}.

\bibitem[{\citenamefont{{Berlin} et~al.}(2015)\citenamefont{{Berlin}, {Gori},
  {Lin}, and {Wang}}}]{2015arXiv150206000B}
\bibinfo{author}{\bibfnamefont{A.}~\bibnamefont{{Berlin}}},
  \bibinfo{author}{\bibfnamefont{S.}~\bibnamefont{{Gori}}},
  \bibinfo{author}{\bibfnamefont{T.}~\bibnamefont{{Lin}}}, \bibnamefont{and}
  \bibinfo{author}{\bibfnamefont{L.-T.} \bibnamefont{{Wang}}},
  \bibinfo{journal}{ArXiv e-prints}  (\bibinfo{year}{2015}),
  \eprint{1502.06000}.

\bibitem[{\citenamefont{{Gherghetta} et~al.}(2015)\citenamefont{{Gherghetta},
  {von Harling}, {Medina}, {Schmidt}, and {Trott}}}]{2015arXiv150207173G}
\bibinfo{author}{\bibfnamefont{T.}~\bibnamefont{{Gherghetta}}},
  \bibinfo{author}{\bibfnamefont{B.}~\bibnamefont{{von Harling}}},
  \bibinfo{author}{\bibfnamefont{A.~D.} \bibnamefont{{Medina}}},
  \bibinfo{author}{\bibfnamefont{M.~A.} \bibnamefont{{Schmidt}}},
  \bibnamefont{and} \bibinfo{author}{\bibfnamefont{T.}~\bibnamefont{{Trott}}},
  \bibinfo{journal}{ArXiv e-prints}  (\bibinfo{year}{2015}),
  \eprint{1502.07173}.

\bibitem[{\citenamefont{{Elor} et~al.}(2015)\citenamefont{{Elor}, {Rodd}, and
  {Slatyer}}}]{2015arXiv150301773E}
\bibinfo{author}{\bibfnamefont{G.}~\bibnamefont{{Elor}}},
  \bibinfo{author}{\bibfnamefont{N.~L.} \bibnamefont{{Rodd}}},
  \bibnamefont{and} \bibinfo{author}{\bibfnamefont{T.~R.}
  \bibnamefont{{Slatyer}}}, \bibinfo{journal}{ArXiv e-prints}
  (\bibinfo{year}{2015}), \eprint{1503.01773}.

\bibitem[{\citenamefont{{Mirabal}}(2013)}]{2013MNRAS.436.2461M}
\bibinfo{author}{\bibfnamefont{N.}~\bibnamefont{{Mirabal}}},
  \bibinfo{journal}{\mnras} \textbf{\bibinfo{volume}{436}},
  \bibinfo{pages}{2461} (\bibinfo{year}{2013}), \eprint{1309.3428}.

\bibitem[{\citenamefont{{Yuan} and {Zhang}}(2014)}]{2014JHEAp...3....1Y}
\bibinfo{author}{\bibfnamefont{Q.}~\bibnamefont{{Yuan}}} \bibnamefont{and}
  \bibinfo{author}{\bibfnamefont{B.}~\bibnamefont{{Zhang}}},
  \bibinfo{journal}{Journal of High Energy Astrophysics}
  \textbf{\bibinfo{volume}{3}}, \bibinfo{pages}{1} (\bibinfo{year}{2014}),
  \eprint{1404.2318}.

\bibitem[{\citenamefont{{Calore} et~al.}(2014)\citenamefont{{Calore}, {Di
  Mauro}, and {Donato}}}]{2014ApJ...796...14C}
\bibinfo{author}{\bibfnamefont{F.}~\bibnamefont{{Calore}}},
  \bibinfo{author}{\bibfnamefont{M.}~\bibnamefont{{Di Mauro}}},
  \bibnamefont{and} \bibinfo{author}{\bibfnamefont{F.}~\bibnamefont{{Donato}}},
  \bibinfo{journal}{\apj} \textbf{\bibinfo{volume}{796}}, \bibinfo{eid}{14}
  (\bibinfo{year}{2014}), \eprint{1406.2706}.

\bibitem[{\citenamefont{{Faucher-Gigu{\`e}re} and
  {Loeb}}(2011)}]{fauchergiguereloeb2011}
\bibinfo{author}{\bibfnamefont{C.-A.} \bibnamefont{{Faucher-Gigu{\`e}re}}}
  \bibnamefont{and} \bibinfo{author}{\bibfnamefont{A.}~\bibnamefont{{Loeb}}},
  \bibinfo{journal}{\mnras} \textbf{\bibinfo{volume}{415}},
  \bibinfo{pages}{3951} (\bibinfo{year}{2011}), \eprint{1012.0573}.

\bibitem[{\citenamefont{{Hooper} et~al.}(2013)\citenamefont{{Hooper}, {Cholis},
  {Linden}, {Siegal-Gaskins}, and {Slatyer}}}]{2013PhRvD..88h3009H}
\bibinfo{author}{\bibfnamefont{D.}~\bibnamefont{{Hooper}}},
  \bibinfo{author}{\bibfnamefont{I.}~\bibnamefont{{Cholis}}},
  \bibinfo{author}{\bibfnamefont{T.}~\bibnamefont{{Linden}}},
  \bibinfo{author}{\bibfnamefont{J.~M.} \bibnamefont{{Siegal-Gaskins}}},
  \bibnamefont{and} \bibinfo{author}{\bibfnamefont{T.~R.}
  \bibnamefont{{Slatyer}}}, \bibinfo{journal}{\prd}
  \textbf{\bibinfo{volume}{88}}, \bibinfo{eid}{083009} (\bibinfo{year}{2013}),
  \eprint{1305.0830}.

\bibitem[{\citenamefont{{Cholis} et~al.}(2014)\citenamefont{{Cholis}, {Hooper},
  and {Linden}}}]{2014arXiv1407.5583C}
\bibinfo{author}{\bibfnamefont{I.}~\bibnamefont{{Cholis}}},
  \bibinfo{author}{\bibfnamefont{D.}~\bibnamefont{{Hooper}}}, \bibnamefont{and}
  \bibinfo{author}{\bibfnamefont{T.}~\bibnamefont{{Linden}}},
  \bibinfo{journal}{ArXiv e-prints}  (\bibinfo{year}{2014}),
  \eprint{1407.5583}.

\bibitem[{\citenamefont{{Aliu} et~al.}(2015)\citenamefont{{Aliu},
  {Archambault}, {Archer}, {Aune}, {Barnacka}, {Beilicke}, {Benbow}, {Bird},
  {Buckley}, {Bugaev} et~al.}}]{2015ApJ...800...61A}
\bibinfo{author}{\bibfnamefont{E.}~\bibnamefont{{Aliu}}},
  \bibinfo{author}{\bibfnamefont{S.}~\bibnamefont{{Archambault}}},
  \bibinfo{author}{\bibfnamefont{A.}~\bibnamefont{{Archer}}},
  \bibinfo{author}{\bibfnamefont{T.}~\bibnamefont{{Aune}}},
  \bibinfo{author}{\bibfnamefont{A.}~\bibnamefont{{Barnacka}}},
  \bibinfo{author}{\bibfnamefont{M.}~\bibnamefont{{Beilicke}}},
  \bibinfo{author}{\bibfnamefont{W.}~\bibnamefont{{Benbow}}},
  \bibinfo{author}{\bibfnamefont{R.}~\bibnamefont{{Bird}}},
  \bibinfo{author}{\bibfnamefont{J.~H.} \bibnamefont{{Buckley}}},
  \bibinfo{author}{\bibfnamefont{V.}~\bibnamefont{{Bugaev}}},
  \bibnamefont{et~al.}, \bibinfo{journal}{\apj} \textbf{\bibinfo{volume}{800}},
  \bibinfo{eid}{61} (\bibinfo{year}{2015}), \eprint{1412.4734}.

\bibitem[{\citenamefont{{Morris} and {Serabyn}}(1996)}]{1996ARA&A..34..645M}
\bibinfo{author}{\bibfnamefont{M.}~\bibnamefont{{Morris}}} \bibnamefont{and}
  \bibinfo{author}{\bibfnamefont{E.}~\bibnamefont{{Serabyn}}},
  \bibinfo{journal}{\araa} \textbf{\bibinfo{volume}{34}}, \bibinfo{pages}{645}
  (\bibinfo{year}{1996}).

\bibitem[{\citenamefont{{Genzel} et~al.}(2010)\citenamefont{{Genzel},
  {Eisenhauer}, and {Gillessen}}}]{genzeletal2010}
\bibinfo{author}{\bibfnamefont{R.}~\bibnamefont{{Genzel}}},
  \bibinfo{author}{\bibfnamefont{F.}~\bibnamefont{{Eisenhauer}}},
  \bibnamefont{and}
  \bibinfo{author}{\bibfnamefont{S.}~\bibnamefont{{Gillessen}}},
  \bibinfo{journal}{Reviews of Modern Physics} \textbf{\bibinfo{volume}{82}},
  \bibinfo{pages}{3121} (\bibinfo{year}{2010}), \eprint{1006.0064}.

\bibitem[{\citenamefont{{Kennicutt} and {Evans}}(2012)}]{2012ARA&A..50..531K}
\bibinfo{author}{\bibfnamefont{R.~C.} \bibnamefont{{Kennicutt}}}
  \bibnamefont{and} \bibinfo{author}{\bibfnamefont{N.~J.}
  \bibnamefont{{Evans}}}, \bibinfo{journal}{\araa}
  \textbf{\bibinfo{volume}{50}}, \bibinfo{pages}{531} (\bibinfo{year}{2012}),
  \eprint{1204.3552}.

\bibitem[{\citenamefont{{Longmore} et~al.}(2013)\citenamefont{{Longmore},
  {Bally}, {Testi}, {Purcell}, {Walsh}, {Bressert}, {Pestalozzi}, {Molinari},
  {Ott}, {Cortese} et~al.}}]{2013MNRAS.429..987L}
\bibinfo{author}{\bibfnamefont{S.~N.} \bibnamefont{{Longmore}}},
  \bibinfo{author}{\bibfnamefont{J.}~\bibnamefont{{Bally}}},
  \bibinfo{author}{\bibfnamefont{L.}~\bibnamefont{{Testi}}},
  \bibinfo{author}{\bibfnamefont{C.~R.} \bibnamefont{{Purcell}}},
  \bibinfo{author}{\bibfnamefont{A.~J.} \bibnamefont{{Walsh}}},
  \bibinfo{author}{\bibfnamefont{E.}~\bibnamefont{{Bressert}}},
  \bibinfo{author}{\bibfnamefont{M.}~\bibnamefont{{Pestalozzi}}},
  \bibinfo{author}{\bibfnamefont{S.}~\bibnamefont{{Molinari}}},
  \bibinfo{author}{\bibfnamefont{J.}~\bibnamefont{{Ott}}},
  \bibinfo{author}{\bibfnamefont{L.}~\bibnamefont{{Cortese}}},
  \bibnamefont{et~al.}, \bibinfo{journal}{\mnras}
  \textbf{\bibinfo{volume}{429}}, \bibinfo{pages}{987} (\bibinfo{year}{2013}),
  \eprint{1208.4256}.

\bibitem[{\citenamefont{{Rosslowe} and {Crowther}}(2015)}]{2015MNRAS.447.2322R}
\bibinfo{author}{\bibfnamefont{C.~K.} \bibnamefont{{Rosslowe}}}
  \bibnamefont{and} \bibinfo{author}{\bibfnamefont{P.~A.}
  \bibnamefont{{Crowther}}}, \bibinfo{journal}{\mnras}
  \textbf{\bibinfo{volume}{447}}, \bibinfo{pages}{2322} (\bibinfo{year}{2015}),
  \eprint{1412.0699}.

\bibitem[{\citenamefont{{Crowther}}(2007)}]{2007ARA&A..45..177C}
\bibinfo{author}{\bibfnamefont{P.~A.} \bibnamefont{{Crowther}}},
  \bibinfo{journal}{\araa} \textbf{\bibinfo{volume}{45}}, \bibinfo{pages}{177}
  (\bibinfo{year}{2007}), \eprint{astro-ph/0610356}.

\bibitem[{\citenamefont{{Mauerhan} et~al.}(2010)\citenamefont{{Mauerhan},
  {Cotera}, {Dong}, {Morris}, {Wang}, {Stolovy}, and
  {Lang}}}]{2010ApJ...725..188M}
\bibinfo{author}{\bibfnamefont{J.~C.} \bibnamefont{{Mauerhan}}},
  \bibinfo{author}{\bibfnamefont{A.}~\bibnamefont{{Cotera}}},
  \bibinfo{author}{\bibfnamefont{H.}~\bibnamefont{{Dong}}},
  \bibinfo{author}{\bibfnamefont{M.~R.} \bibnamefont{{Morris}}},
  \bibinfo{author}{\bibfnamefont{Q.~D.} \bibnamefont{{Wang}}},
  \bibinfo{author}{\bibfnamefont{S.~R.} \bibnamefont{{Stolovy}}},
  \bibnamefont{and} \bibinfo{author}{\bibfnamefont{C.}~\bibnamefont{{Lang}}},
  \bibinfo{journal}{\apj} \textbf{\bibinfo{volume}{725}}, \bibinfo{eid}{188}
  (\bibinfo{year}{2010}), \eprint{1009.2769}.

\bibitem[{\citenamefont{{Dong} et~al.}(2012)\citenamefont{{Dong}, {Wang}, and
  {Morris}}}]{2012MNRAS.425..884D}
\bibinfo{author}{\bibfnamefont{H.}~\bibnamefont{{Dong}}},
  \bibinfo{author}{\bibfnamefont{Q.~D.} \bibnamefont{{Wang}}},
  \bibnamefont{and} \bibinfo{author}{\bibfnamefont{M.~R.}
  \bibnamefont{{Morris}}}, \bibinfo{journal}{\mnras}
  \textbf{\bibinfo{volume}{425}}, \bibinfo{pages}{884} (\bibinfo{year}{2012}),
  \eprint{1204.6298}.

\bibitem[{\citenamefont{{Kanarek} et~al.}(2014)\citenamefont{{Kanarek},
  {Shara}, {Faherty}, {Zurek}, and {Moffat}}}]{2014arXiv1403.0975K}
\bibinfo{author}{\bibfnamefont{G.~C.} \bibnamefont{{Kanarek}}},
  \bibinfo{author}{\bibfnamefont{M.~M.} \bibnamefont{{Shara}}},
  \bibinfo{author}{\bibfnamefont{J.~K.} \bibnamefont{{Faherty}}},
  \bibinfo{author}{\bibfnamefont{D.}~\bibnamefont{{Zurek}}}, \bibnamefont{and}
  \bibinfo{author}{\bibfnamefont{A.~F.~J.} \bibnamefont{{Moffat}}},
  \bibinfo{journal}{ArXiv e-prints}  (\bibinfo{year}{2014}),
  \eprint{1403.0975}.

\bibitem[{\citenamefont{{Kruijssen} et~al.}(2014)\citenamefont{{Kruijssen},
  {Longmore}, {Elmegreen}, {Murray}, {Bally}, {Testi}, and
  {Kennicutt}}}]{2014MNRAS.440.3370K}
\bibinfo{author}{\bibfnamefont{J.~M.~D.} \bibnamefont{{Kruijssen}}},
  \bibinfo{author}{\bibfnamefont{S.~N.} \bibnamefont{{Longmore}}},
  \bibinfo{author}{\bibfnamefont{B.~G.} \bibnamefont{{Elmegreen}}},
  \bibinfo{author}{\bibfnamefont{N.}~\bibnamefont{{Murray}}},
  \bibinfo{author}{\bibfnamefont{J.}~\bibnamefont{{Bally}}},
  \bibinfo{author}{\bibfnamefont{L.}~\bibnamefont{{Testi}}}, \bibnamefont{and}
  \bibinfo{author}{\bibfnamefont{R.~C.} \bibnamefont{{Kennicutt}}},
  \bibinfo{journal}{\mnras} \textbf{\bibinfo{volume}{440}},
  \bibinfo{pages}{3370} (\bibinfo{year}{2014}), \eprint{1303.6286}.

\bibitem[{\citenamefont{{Cirelli} et~al.}(2011)\citenamefont{{Cirelli},
  {Corcella}, {Hektor}, {H{\"u}tsi}, {Kadastik}, {Panci}, {Raidal}, {Sala}, and
  {Strumia}}}]{2011JCAP...03..051C}
\bibinfo{author}{\bibfnamefont{M.}~\bibnamefont{{Cirelli}}},
  \bibinfo{author}{\bibfnamefont{G.}~\bibnamefont{{Corcella}}},
  \bibinfo{author}{\bibfnamefont{A.}~\bibnamefont{{Hektor}}},
  \bibinfo{author}{\bibfnamefont{G.}~\bibnamefont{{H{\"u}tsi}}},
  \bibinfo{author}{\bibfnamefont{M.}~\bibnamefont{{Kadastik}}},
  \bibinfo{author}{\bibfnamefont{P.}~\bibnamefont{{Panci}}},
  \bibinfo{author}{\bibfnamefont{M.}~\bibnamefont{{Raidal}}},
  \bibinfo{author}{\bibfnamefont{F.}~\bibnamefont{{Sala}}}, \bibnamefont{and}
  \bibinfo{author}{\bibfnamefont{A.}~\bibnamefont{{Strumia}}},
  \bibinfo{journal}{\jcap} \textbf{\bibinfo{volume}{3}}, \bibinfo{eid}{051}
  (\bibinfo{year}{2011}), \eprint{1012.4515}.

\bibitem[{\citenamefont{{Pierbattista}
  et~al.}(2015)\citenamefont{{Pierbattista}, {Harding}, {Grenier}, {Johnson},
  {Caraveo}, {Kerr}, and {Gonthier}}}]{2015A&A...575A...3P}
\bibinfo{author}{\bibfnamefont{M.}~\bibnamefont{{Pierbattista}}},
  \bibinfo{author}{\bibfnamefont{A.~K.} \bibnamefont{{Harding}}},
  \bibinfo{author}{\bibfnamefont{I.~A.} \bibnamefont{{Grenier}}},
  \bibinfo{author}{\bibfnamefont{T.~J.} \bibnamefont{{Johnson}}},
  \bibinfo{author}{\bibfnamefont{P.~A.} \bibnamefont{{Caraveo}}},
  \bibinfo{author}{\bibfnamefont{M.}~\bibnamefont{{Kerr}}}, \bibnamefont{and}
  \bibinfo{author}{\bibfnamefont{P.~L.} \bibnamefont{{Gonthier}}},
  \bibinfo{journal}{\aap} \textbf{\bibinfo{volume}{575}}, \bibinfo{eid}{A3}
  (\bibinfo{year}{2015}), \eprint{1403.3849}.

\bibitem[{\citenamefont{{Yusifov} and {K{\"u}{\c
  c}{\"u}k}}(2004)}]{2004A&A...422..545Y}
\bibinfo{author}{\bibfnamefont{I.}~\bibnamefont{{Yusifov}}} \bibnamefont{and}
  \bibinfo{author}{\bibfnamefont{I.}~\bibnamefont{{K{\"u}{\c c}{\"u}k}}},
  \bibinfo{journal}{\aap} \textbf{\bibinfo{volume}{422}}, \bibinfo{pages}{545}
  (\bibinfo{year}{2004}), \eprint{astro-ph/0405559}.

\bibitem[{\citenamefont{{Manchester} et~al.}(2005)\citenamefont{{Manchester},
  {Hobbs}, {Teoh}, and {Hobbs}}}]{manchesteretal2005}
\bibinfo{author}{\bibfnamefont{R.~N.} \bibnamefont{{Manchester}}},
  \bibinfo{author}{\bibfnamefont{G.~B.} \bibnamefont{{Hobbs}}},
  \bibinfo{author}{\bibfnamefont{A.}~\bibnamefont{{Teoh}}}, \bibnamefont{and}
  \bibinfo{author}{\bibfnamefont{M.}~\bibnamefont{{Hobbs}}},
  \bibinfo{journal}{\aj} \textbf{\bibinfo{volume}{129}}, \bibinfo{pages}{1993}
  (\bibinfo{year}{2005}), \eprint{arXiv:astro-ph/0412641}.

\bibitem[{\citenamefont{{Carlson} and {Profumo}}(2014)}]{2014PhRvD..90b3015C}
\bibinfo{author}{\bibfnamefont{E.}~\bibnamefont{{Carlson}}} \bibnamefont{and}
  \bibinfo{author}{\bibfnamefont{S.}~\bibnamefont{{Profumo}}},
  \bibinfo{journal}{\prd} \textbf{\bibinfo{volume}{90}}, \bibinfo{eid}{023015}
  (\bibinfo{year}{2014}), \eprint{1405.7685}.

\bibitem[{\citenamefont{{Petrovi{\'c}}
  et~al.}(2014)\citenamefont{{Petrovi{\'c}}, {Dario Serpico}, and {Zaharija{\v
  s}}}}]{2014JCAP...10..052P}
\bibinfo{author}{\bibfnamefont{J.}~\bibnamefont{{Petrovi{\'c}}}},
  \bibinfo{author}{\bibfnamefont{P.}~\bibnamefont{{Dario Serpico}}},
  \bibnamefont{and}
  \bibinfo{author}{\bibfnamefont{G.}~\bibnamefont{{Zaharija{\v s}}}},
  \bibinfo{journal}{\jcap} \textbf{\bibinfo{volume}{10}}, \bibinfo{eid}{052}
  (\bibinfo{year}{2014}), \eprint{1405.7928}.

\bibitem[{\citenamefont{{Kistler} et~al.}(2015)\citenamefont{{Kistler},
  {O'Leary}, {Dexter}, and {Kerr}}}]{2015toappear}
\bibinfo{author}{\bibfnamefont{M.~D.} \bibnamefont{{Kistler}}},
  \bibinfo{author}{\bibfnamefont{R.~M.} \bibnamefont{{O'Leary}}},
  \bibinfo{author}{\bibfnamefont{J.}~\bibnamefont{{Dexter}}}, \bibnamefont{and}
  \bibinfo{author}{\bibfnamefont{M.}~\bibnamefont{{Kerr}}}
  (\bibinfo{year}{2015}), \bibinfo{note}{to appear}.

\bibitem[{\citenamefont{{Faucher-Gigu{\`e}re} and
  {Kaspi}}(2006)}]{fauchergiguerekaspi2006}
\bibinfo{author}{\bibfnamefont{C.-A.} \bibnamefont{{Faucher-Gigu{\`e}re}}}
  \bibnamefont{and} \bibinfo{author}{\bibfnamefont{V.~M.}
  \bibnamefont{{Kaspi}}}, \bibinfo{journal}{\apj}
  \textbf{\bibinfo{volume}{643}}, \bibinfo{pages}{332} (\bibinfo{year}{2006}),
  \eprint{arXiv:astro-ph/0512585}.

\bibitem[{\citenamefont{{Dexter} and {O'Leary}}(2014)}]{dexter2014}
\bibinfo{author}{\bibfnamefont{J.}~\bibnamefont{{Dexter}}} \bibnamefont{and}
  \bibinfo{author}{\bibfnamefont{R.~M.} \bibnamefont{{O'Leary}}},
  \bibinfo{journal}{\apjl} \textbf{\bibinfo{volume}{783}}, \bibinfo{eid}{L7}
  (\bibinfo{year}{2014}), \eprint{1310.7022}.

\bibitem[{\citenamefont{{Li} et~al.}(2011)\citenamefont{{Li}, {Chornock},
  {Leaman}, {Filippenko}, {Poznanski}, {Wang}, {Ganeshalingam}, and
  {Mannucci}}}]{2011MNRAS.412.1473L}
\bibinfo{author}{\bibfnamefont{W.}~\bibnamefont{{Li}}},
  \bibinfo{author}{\bibfnamefont{R.}~\bibnamefont{{Chornock}}},
  \bibinfo{author}{\bibfnamefont{J.}~\bibnamefont{{Leaman}}},
  \bibinfo{author}{\bibfnamefont{A.~V.} \bibnamefont{{Filippenko}}},
  \bibinfo{author}{\bibfnamefont{D.}~\bibnamefont{{Poznanski}}},
  \bibinfo{author}{\bibfnamefont{X.}~\bibnamefont{{Wang}}},
  \bibinfo{author}{\bibfnamefont{M.}~\bibnamefont{{Ganeshalingam}}},
  \bibnamefont{and}
  \bibinfo{author}{\bibfnamefont{F.}~\bibnamefont{{Mannucci}}},
  \bibinfo{journal}{\mnras} \textbf{\bibinfo{volume}{412}},
  \bibinfo{pages}{1473} (\bibinfo{year}{2011}), \eprint{1006.4613}.

\bibitem[{\citenamefont{{Groh} et~al.}(2013)\citenamefont{{Groh}, {Meynet},
  {Georgy}, and {Ekstr{\"o}m}}}]{2013A&A...558A.131G}
\bibinfo{author}{\bibfnamefont{J.~H.} \bibnamefont{{Groh}}},
  \bibinfo{author}{\bibfnamefont{G.}~\bibnamefont{{Meynet}}},
  \bibinfo{author}{\bibfnamefont{C.}~\bibnamefont{{Georgy}}}, \bibnamefont{and}
  \bibinfo{author}{\bibfnamefont{S.}~\bibnamefont{{Ekstr{\"o}m}}},
  \bibinfo{journal}{\aap} \textbf{\bibinfo{volume}{558}}, \bibinfo{eid}{A131}
  (\bibinfo{year}{2013}), \eprint{1308.4681}.

\bibitem[{\citenamefont{{Hobbs} et~al.}(2005)\citenamefont{{Hobbs}, {Lorimer},
  {Lyne}, and {Kramer}}}]{2005MNRAS.360..974H}
\bibinfo{author}{\bibfnamefont{G.}~\bibnamefont{{Hobbs}}},
  \bibinfo{author}{\bibfnamefont{D.~R.} \bibnamefont{{Lorimer}}},
  \bibinfo{author}{\bibfnamefont{A.~G.} \bibnamefont{{Lyne}}},
  \bibnamefont{and} \bibinfo{author}{\bibfnamefont{M.}~\bibnamefont{{Kramer}}},
  \bibinfo{journal}{\mnras} \textbf{\bibinfo{volume}{360}},
  \bibinfo{pages}{974} (\bibinfo{year}{2005}), \eprint{astro-ph/0504584}.

\bibitem[{\citenamefont{{Bovy}}(2014)}]{bovy2015}
\bibinfo{author}{\bibfnamefont{J.}~\bibnamefont{{Bovy}}},
  \bibinfo{journal}{ArXiv e-prints}  (\bibinfo{year}{2014}),
  \eprint{1412.3451}.

\bibitem[{\citenamefont{{Hou} et~al.}(2014)\citenamefont{{Hou}, {Smith},
  {Guillemot}, {Cheung}, {Cognard}, {Craig}, {Espinoza}, {Johnston}, {Kramer},
  {Reimer} et~al.}}]{2014A&A...570A..44H}
\bibinfo{author}{\bibfnamefont{X.}~\bibnamefont{{Hou}}},
  \bibinfo{author}{\bibfnamefont{D.~A.} \bibnamefont{{Smith}}},
  \bibinfo{author}{\bibfnamefont{L.}~\bibnamefont{{Guillemot}}},
  \bibinfo{author}{\bibfnamefont{C.~C.} \bibnamefont{{Cheung}}},
  \bibinfo{author}{\bibfnamefont{I.}~\bibnamefont{{Cognard}}},
  \bibinfo{author}{\bibfnamefont{H.~A.} \bibnamefont{{Craig}}},
  \bibinfo{author}{\bibfnamefont{C.~M.} \bibnamefont{{Espinoza}}},
  \bibinfo{author}{\bibfnamefont{S.}~\bibnamefont{{Johnston}}},
  \bibinfo{author}{\bibfnamefont{M.}~\bibnamefont{{Kramer}}},
  \bibinfo{author}{\bibfnamefont{O.}~\bibnamefont{{Reimer}}},
  \bibnamefont{et~al.}, \bibinfo{journal}{\aap} \textbf{\bibinfo{volume}{570}},
  \bibinfo{eid}{A44} (\bibinfo{year}{2014}), \eprint{1407.6271}.

\bibitem[{\citenamefont{{Nolan} et~al.}(2012)\citenamefont{{Nolan}, {Abdo},
  {Ackermann}, {Ajello}, {Allafort}, {Antolini}, {Atwood}, {Axelsson},
  {Baldini}, {Ballet} et~al.}}]{2012ApJS..199...31N}
\bibinfo{author}{\bibfnamefont{P.~L.} \bibnamefont{{Nolan}}},
  \bibinfo{author}{\bibfnamefont{A.~A.} \bibnamefont{{Abdo}}},
  \bibinfo{author}{\bibfnamefont{M.}~\bibnamefont{{Ackermann}}},
  \bibinfo{author}{\bibfnamefont{M.}~\bibnamefont{{Ajello}}},
  \bibinfo{author}{\bibfnamefont{A.}~\bibnamefont{{Allafort}}},
  \bibinfo{author}{\bibfnamefont{E.}~\bibnamefont{{Antolini}}},
  \bibinfo{author}{\bibfnamefont{W.~B.} \bibnamefont{{Atwood}}},
  \bibinfo{author}{\bibfnamefont{M.}~\bibnamefont{{Axelsson}}},
  \bibinfo{author}{\bibfnamefont{L.}~\bibnamefont{{Baldini}}},
  \bibinfo{author}{\bibfnamefont{J.}~\bibnamefont{{Ballet}}},
  \bibnamefont{et~al.}, \bibinfo{journal}{\apjs}
  \textbf{\bibinfo{volume}{199}}, \bibinfo{eid}{31} (\bibinfo{year}{2012}),
  \eprint{1108.1435}.

\bibitem[{\citenamefont{{Watters} et~al.}(2009)\citenamefont{{Watters},
  {Romani}, {Weltevrede}, and {Johnston}}}]{2009ApJ...695.1289W}
\bibinfo{author}{\bibfnamefont{K.~P.} \bibnamefont{{Watters}}},
  \bibinfo{author}{\bibfnamefont{R.~W.} \bibnamefont{{Romani}}},
  \bibinfo{author}{\bibfnamefont{P.}~\bibnamefont{{Weltevrede}}},
  \bibnamefont{and}
  \bibinfo{author}{\bibfnamefont{S.}~\bibnamefont{{Johnston}}},
  \bibinfo{journal}{\apj} \textbf{\bibinfo{volume}{695}}, \bibinfo{pages}{1289}
  (\bibinfo{year}{2009}), \eprint{0812.3931}.

\bibitem[{\citenamefont{{The Fermi-LAT
  Collaboration}}(2015)}]{2015arXiv150102003T}
\bibinfo{author}{\bibnamefont{{The Fermi-LAT Collaboration}}},
  \bibinfo{journal}{ArXiv e-prints}  (\bibinfo{year}{2015}),
  \eprint{1501.02003}.

\bibitem[{\citenamefont{{Arons}}(1996)}]{1996A&AS..120C..49A}
\bibinfo{author}{\bibfnamefont{J.}~\bibnamefont{{Arons}}},
  \bibinfo{journal}{\aaps} \textbf{\bibinfo{volume}{120}}, \bibinfo{pages}{C49}
  (\bibinfo{year}{1996}).

\bibitem[{\citenamefont{{Wang} and {Hirotani}}(2011)}]{2011ApJ...736..127W}
\bibinfo{author}{\bibfnamefont{R.-B.} \bibnamefont{{Wang}}} \bibnamefont{and}
  \bibinfo{author}{\bibfnamefont{K.}~\bibnamefont{{Hirotani}}},
  \bibinfo{journal}{\apj} \textbf{\bibinfo{volume}{736}}, \bibinfo{eid}{127}
  (\bibinfo{year}{2011}), \eprint{1105.3030}.

\bibitem[{\citenamefont{{Ackermann} et~al.}(2013)\citenamefont{{Ackermann},
  {Ajello}, {Allafort}, {Asano}, {Atwood}, {Baldini}, {Ballet}, {Barbiellini},
  {Bastieri}, {Bechtol} et~al.}}]{2013fermipsf}
\bibinfo{author}{\bibfnamefont{M.}~\bibnamefont{{Ackermann}}},
  \bibinfo{author}{\bibfnamefont{M.}~\bibnamefont{{Ajello}}},
  \bibinfo{author}{\bibfnamefont{A.}~\bibnamefont{{Allafort}}},
  \bibinfo{author}{\bibfnamefont{K.}~\bibnamefont{{Asano}}},
  \bibinfo{author}{\bibfnamefont{W.~B.} \bibnamefont{{Atwood}}},
  \bibinfo{author}{\bibfnamefont{L.}~\bibnamefont{{Baldini}}},
  \bibinfo{author}{\bibfnamefont{J.}~\bibnamefont{{Ballet}}},
  \bibinfo{author}{\bibfnamefont{G.}~\bibnamefont{{Barbiellini}}},
  \bibinfo{author}{\bibfnamefont{D.}~\bibnamefont{{Bastieri}}},
  \bibinfo{author}{\bibfnamefont{K.}~\bibnamefont{{Bechtol}}},
  \bibnamefont{et~al.}, \bibinfo{journal}{\apj} \textbf{\bibinfo{volume}{765}},
  \bibinfo{eid}{54} (\bibinfo{year}{2013}).

\bibitem[{\citenamefont{{Romani} et~al.}(2011)\citenamefont{{Romani}, {Kerr},
  {Craig}, {Johnston}, {Cognard}, and {Smith}}}]{2011ApJ...738..114R}
\bibinfo{author}{\bibfnamefont{R.~W.} \bibnamefont{{Romani}}},
  \bibinfo{author}{\bibfnamefont{M.}~\bibnamefont{{Kerr}}},
  \bibinfo{author}{\bibfnamefont{H.~A.} \bibnamefont{{Craig}}},
  \bibinfo{author}{\bibfnamefont{S.}~\bibnamefont{{Johnston}}},
  \bibinfo{author}{\bibfnamefont{I.}~\bibnamefont{{Cognard}}},
  \bibnamefont{and} \bibinfo{author}{\bibfnamefont{D.~A.}
  \bibnamefont{{Smith}}}, \bibinfo{journal}{\apj}
  \textbf{\bibinfo{volume}{738}}, \bibinfo{eid}{114} (\bibinfo{year}{2011}),
  \eprint{1106.5762}.

\bibitem[{\citenamefont{{Crocker} et~al.}(2011)\citenamefont{{Crocker},
  {Jones}, {Aharonian}, {Law}, {Melia}, {Oka}, and
  {Ott}}}]{2011MNRAS.413..763C}
\bibinfo{author}{\bibfnamefont{R.~M.} \bibnamefont{{Crocker}}},
  \bibinfo{author}{\bibfnamefont{D.~I.} \bibnamefont{{Jones}}},
  \bibinfo{author}{\bibfnamefont{F.}~\bibnamefont{{Aharonian}}},
  \bibinfo{author}{\bibfnamefont{C.~J.} \bibnamefont{{Law}}},
  \bibinfo{author}{\bibfnamefont{F.}~\bibnamefont{{Melia}}},
  \bibinfo{author}{\bibfnamefont{T.}~\bibnamefont{{Oka}}}, \bibnamefont{and}
  \bibinfo{author}{\bibfnamefont{J.}~\bibnamefont{{Ott}}},
  \bibinfo{journal}{\mnras} \textbf{\bibinfo{volume}{413}},
  \bibinfo{pages}{763} (\bibinfo{year}{2011}), \eprint{1011.0206}.

\bibitem[{\citenamefont{{Lacki}}(2014)}]{2014MNRAS.444L..39L}
\bibinfo{author}{\bibfnamefont{B.~C.} \bibnamefont{{Lacki}}},
  \bibinfo{journal}{\mnras} \textbf{\bibinfo{volume}{444}},
  \bibinfo{pages}{L39} (\bibinfo{year}{2014}), \eprint{1304.6137}.

\bibitem[{\citenamefont{{Ruderman} and
  {Sutherland}}(1975)}]{1975ApJ...196...51R}
\bibinfo{author}{\bibfnamefont{M.~A.} \bibnamefont{{Ruderman}}}
  \bibnamefont{and} \bibinfo{author}{\bibfnamefont{P.~G.}
  \bibnamefont{{Sutherland}}}, \bibinfo{journal}{\apj}
  \textbf{\bibinfo{volume}{196}}, \bibinfo{pages}{51} (\bibinfo{year}{1975}).

\bibitem[{\citenamefont{{Arons}}(1983)}]{1983ApJ...266..215A}
\bibinfo{author}{\bibfnamefont{J.}~\bibnamefont{{Arons}}},
  \bibinfo{journal}{\apj} \textbf{\bibinfo{volume}{266}}, \bibinfo{pages}{215}
  (\bibinfo{year}{1983}).

\bibitem[{\citenamefont{{Cheng} et~al.}(1986)\citenamefont{{Cheng}, {Ho}, and
  {Ruderman}}}]{1986ApJ...300..500C}
\bibinfo{author}{\bibfnamefont{K.~S.} \bibnamefont{{Cheng}}},
  \bibinfo{author}{\bibfnamefont{C.}~\bibnamefont{{Ho}}}, \bibnamefont{and}
  \bibinfo{author}{\bibfnamefont{M.}~\bibnamefont{{Ruderman}}},
  \bibinfo{journal}{\apj} \textbf{\bibinfo{volume}{300}}, \bibinfo{pages}{500}
  (\bibinfo{year}{1986}).

\bibitem[{\citenamefont{{Romani}}(1996)}]{1996ApJ...470..469R}
\bibinfo{author}{\bibfnamefont{R.~W.} \bibnamefont{{Romani}}},
  \bibinfo{journal}{\apj} \textbf{\bibinfo{volume}{470}}, \bibinfo{pages}{469}
  (\bibinfo{year}{1996}).

\bibitem[{\citenamefont{{Harding} et~al.}(2008)\citenamefont{{Harding},
  {Stern}, {Dyks}, and {Frackowiak}}}]{2008ApJ...680.1378H}
\bibinfo{author}{\bibfnamefont{A.~K.} \bibnamefont{{Harding}}},
  \bibinfo{author}{\bibfnamefont{J.~V.} \bibnamefont{{Stern}}},
  \bibinfo{author}{\bibfnamefont{J.}~\bibnamefont{{Dyks}}}, \bibnamefont{and}
  \bibinfo{author}{\bibfnamefont{M.}~\bibnamefont{{Frackowiak}}},
  \bibinfo{journal}{\apj} \textbf{\bibinfo{volume}{680}}, \bibinfo{pages}{1378}
  (\bibinfo{year}{2008}), \eprint{0803.0699}.

\bibitem[{\citenamefont{{Bai} and {Spitkovsky}}(2010)}]{2010ApJ...715.1282B}
\bibinfo{author}{\bibfnamefont{X.-N.} \bibnamefont{{Bai}}} \bibnamefont{and}
  \bibinfo{author}{\bibfnamefont{A.}~\bibnamefont{{Spitkovsky}}},
  \bibinfo{journal}{\apj} \textbf{\bibinfo{volume}{715}}, \bibinfo{pages}{1282}
  (\bibinfo{year}{2010}), \eprint{0910.5741}.

\bibitem[{\citenamefont{{Kalapotharakos}
  et~al.}(2014)\citenamefont{{Kalapotharakos}, {Harding}, and
  {Kazanas}}}]{2014ApJ...793...97K}
\bibinfo{author}{\bibfnamefont{C.}~\bibnamefont{{Kalapotharakos}}},
  \bibinfo{author}{\bibfnamefont{A.~K.} \bibnamefont{{Harding}}},
  \bibnamefont{and}
  \bibinfo{author}{\bibfnamefont{D.}~\bibnamefont{{Kazanas}}},
  \bibinfo{journal}{\apj} \textbf{\bibinfo{volume}{793}}, \bibinfo{eid}{97}
  (\bibinfo{year}{2014}), \eprint{1310.3545}.

\bibitem[{\citenamefont{{Chen} and {Beloborodov}}(2014)}]{2014ApJ...795L..22C}
\bibinfo{author}{\bibfnamefont{A.~Y.} \bibnamefont{{Chen}}} \bibnamefont{and}
  \bibinfo{author}{\bibfnamefont{A.~M.} \bibnamefont{{Beloborodov}}},
  \bibinfo{journal}{\apjl} \textbf{\bibinfo{volume}{795}}, \bibinfo{eid}{L22}
  (\bibinfo{year}{2014}), \eprint{1406.7834}.

\bibitem[{\citenamefont{{Philippov} et~al.}(2014)\citenamefont{{Philippov},
  {Spitkovsky}, and {Cerutti}}}]{2014arXiv1412.0673P}
\bibinfo{author}{\bibfnamefont{A.~A.} \bibnamefont{{Philippov}}},
  \bibinfo{author}{\bibfnamefont{A.}~\bibnamefont{{Spitkovsky}}},
  \bibnamefont{and}
  \bibinfo{author}{\bibfnamefont{B.}~\bibnamefont{{Cerutti}}},
  \bibinfo{journal}{ArXiv e-prints}  (\bibinfo{year}{2014}),
  \eprint{1412.0673}.

\bibitem[{\citenamefont{{Nordhaus} et~al.}(2012)\citenamefont{{Nordhaus},
  {Brandt}, {Burrows}, and {Almgren}}}]{2012MNRAS.423.1805N}
\bibinfo{author}{\bibfnamefont{J.}~\bibnamefont{{Nordhaus}}},
  \bibinfo{author}{\bibfnamefont{T.~D.} \bibnamefont{{Brandt}}},
  \bibinfo{author}{\bibfnamefont{A.}~\bibnamefont{{Burrows}}},
  \bibnamefont{and}
  \bibinfo{author}{\bibfnamefont{A.}~\bibnamefont{{Almgren}}},
  \bibinfo{journal}{\mnras} \textbf{\bibinfo{volume}{423}},
  \bibinfo{pages}{1805} (\bibinfo{year}{2012}), \eprint{1112.3342}.

\bibitem[{\citenamefont{{Ott} et~al.}(2013)\citenamefont{{Ott}, {Abdikamalov},
  {M{\"o}sta}, {Haas}, {Drasco}, {O'Connor}, {Reisswig}, {Meakin}, and
  {Schnetter}}}]{2013ApJ...768..115O}
\bibinfo{author}{\bibfnamefont{C.~D.} \bibnamefont{{Ott}}},
  \bibinfo{author}{\bibfnamefont{E.}~\bibnamefont{{Abdikamalov}}},
  \bibinfo{author}{\bibfnamefont{P.}~\bibnamefont{{M{\"o}sta}}},
  \bibinfo{author}{\bibfnamefont{R.}~\bibnamefont{{Haas}}},
  \bibinfo{author}{\bibfnamefont{S.}~\bibnamefont{{Drasco}}},
  \bibinfo{author}{\bibfnamefont{E.~P.} \bibnamefont{{O'Connor}}},
  \bibinfo{author}{\bibfnamefont{C.}~\bibnamefont{{Reisswig}}},
  \bibinfo{author}{\bibfnamefont{C.~A.} \bibnamefont{{Meakin}}},
  \bibnamefont{and}
  \bibinfo{author}{\bibfnamefont{E.}~\bibnamefont{{Schnetter}}},
  \bibinfo{journal}{\apj} \textbf{\bibinfo{volume}{768}}, \bibinfo{eid}{115}
  (\bibinfo{year}{2013}), \eprint{1210.6674}.

\bibitem[{\citenamefont{{Wongwathanarat}
  et~al.}(2013)\citenamefont{{Wongwathanarat}, {Janka}, and
  {M{\"u}ller}}}]{2013A&A...552A.126W}
\bibinfo{author}{\bibfnamefont{A.}~\bibnamefont{{Wongwathanarat}}},
  \bibinfo{author}{\bibfnamefont{H.-T.} \bibnamefont{{Janka}}},
  \bibnamefont{and}
  \bibinfo{author}{\bibfnamefont{E.}~\bibnamefont{{M{\"u}ller}}},
  \bibinfo{journal}{\aap} \textbf{\bibinfo{volume}{552}}, \bibinfo{eid}{A126}
  (\bibinfo{year}{2013}), \eprint{1210.8148}.

\bibitem[{\citenamefont{{Bruenn} et~al.}(2014)\citenamefont{{Bruenn}, {Lentz},
  {Hix}, {Mezzacappa}, {Harris}, {Bronson Messer}, {Endeve}, {Blondin},
  {Chertkow}, {Lingerfelt} et~al.}}]{2014arXiv1409.5779B}
\bibinfo{author}{\bibfnamefont{S.~W.} \bibnamefont{{Bruenn}}},
  \bibinfo{author}{\bibfnamefont{E.~J.} \bibnamefont{{Lentz}}},
  \bibinfo{author}{\bibfnamefont{W.~R.} \bibnamefont{{Hix}}},
  \bibinfo{author}{\bibfnamefont{A.}~\bibnamefont{{Mezzacappa}}},
  \bibinfo{author}{\bibfnamefont{J.~A.} \bibnamefont{{Harris}}},
  \bibinfo{author}{\bibfnamefont{O.~E.} \bibnamefont{{Bronson Messer}}},
  \bibinfo{author}{\bibfnamefont{E.}~\bibnamefont{{Endeve}}},
  \bibinfo{author}{\bibfnamefont{J.~M.} \bibnamefont{{Blondin}}},
  \bibinfo{author}{\bibfnamefont{M.~A.} \bibnamefont{{Chertkow}}},
  \bibinfo{author}{\bibfnamefont{E.~J.} \bibnamefont{{Lingerfelt}}},
  \bibnamefont{et~al.}, \bibinfo{journal}{ArXiv e-prints}
  (\bibinfo{year}{2014}), \eprint{1409.5779}.

\bibitem[{\citenamefont{{Ackermann} et~al.}(2012)\citenamefont{{Ackermann},
  {Ajello}, {Atwood}, {Baldini}, {Ballet}, {Barbiellini}, {Bastieri},
  {Bechtol}, {Bellazzini}, {Berenji} et~al.}}]{2012ApJ...750....3A}
\bibinfo{author}{\bibfnamefont{M.}~\bibnamefont{{Ackermann}}},
  \bibinfo{author}{\bibfnamefont{M.}~\bibnamefont{{Ajello}}},
  \bibinfo{author}{\bibfnamefont{W.~B.} \bibnamefont{{Atwood}}},
  \bibinfo{author}{\bibfnamefont{L.}~\bibnamefont{{Baldini}}},
  \bibinfo{author}{\bibfnamefont{J.}~\bibnamefont{{Ballet}}},
  \bibinfo{author}{\bibfnamefont{G.}~\bibnamefont{{Barbiellini}}},
  \bibinfo{author}{\bibfnamefont{D.}~\bibnamefont{{Bastieri}}},
  \bibinfo{author}{\bibfnamefont{K.}~\bibnamefont{{Bechtol}}},
  \bibinfo{author}{\bibfnamefont{R.}~\bibnamefont{{Bellazzini}}},
  \bibinfo{author}{\bibfnamefont{B.}~\bibnamefont{{Berenji}}},
  \bibnamefont{et~al.}, \bibinfo{journal}{\apj} \textbf{\bibinfo{volume}{750}},
  \bibinfo{eid}{3} (\bibinfo{year}{2012}), \eprint{1202.4039}.

\bibitem[{\citenamefont{{Abdo} et~al.}(2009)\citenamefont{{Abdo}, {Allen},
  {Aune}, {Berley}, {Chen}, {Christopher}, {DeYoung}, {Dingus}, {Ellsworth},
  {Gonzalez} et~al.}}]{2009ApJ...700L.127A}
\bibinfo{author}{\bibfnamefont{A.~A.} \bibnamefont{{Abdo}}},
  \bibinfo{author}{\bibfnamefont{B.~T.} \bibnamefont{{Allen}}},
  \bibinfo{author}{\bibfnamefont{T.}~\bibnamefont{{Aune}}},
  \bibinfo{author}{\bibfnamefont{D.}~\bibnamefont{{Berley}}},
  \bibinfo{author}{\bibfnamefont{C.}~\bibnamefont{{Chen}}},
  \bibinfo{author}{\bibfnamefont{G.~E.} \bibnamefont{{Christopher}}},
  \bibinfo{author}{\bibfnamefont{T.}~\bibnamefont{{DeYoung}}},
  \bibinfo{author}{\bibfnamefont{B.~L.} \bibnamefont{{Dingus}}},
  \bibinfo{author}{\bibfnamefont{R.~W.} \bibnamefont{{Ellsworth}}},
  \bibinfo{author}{\bibfnamefont{M.~M.} \bibnamefont{{Gonzalez}}},
  \bibnamefont{et~al.}, \bibinfo{journal}{\apjl}
  \textbf{\bibinfo{volume}{700}}, \bibinfo{pages}{L127} (\bibinfo{year}{2009}),
  \eprint{0904.1018}.

\bibitem[{\citenamefont{{Y{\"u}ksel} et~al.}(2009)\citenamefont{{Y{\"u}ksel},
  {Kistler}, and {Stanev}}}]{yuksel2009}
\bibinfo{author}{\bibfnamefont{H.}~\bibnamefont{{Y{\"u}ksel}}},
  \bibinfo{author}{\bibfnamefont{M.~D.} \bibnamefont{{Kistler}}},
  \bibnamefont{and} \bibinfo{author}{\bibfnamefont{T.}~\bibnamefont{{Stanev}}},
  \bibinfo{journal}{Physical Review Letters} \textbf{\bibinfo{volume}{103}},
  \bibinfo{eid}{051101} (\bibinfo{year}{2009}), \eprint{0810.2784}.

\bibitem[{\citenamefont{{Geringer-Sameth}
  et~al.}(2014)\citenamefont{{Geringer-Sameth}, {Koushiappas}, and
  {Walker}}}]{2014arXiv1410.2242G}
\bibinfo{author}{\bibfnamefont{A.}~\bibnamefont{{Geringer-Sameth}}},
  \bibinfo{author}{\bibfnamefont{S.~M.} \bibnamefont{{Koushiappas}}},
  \bibnamefont{and} \bibinfo{author}{\bibfnamefont{M.~G.}
  \bibnamefont{{Walker}}}, \bibinfo{journal}{ArXiv e-prints}
  (\bibinfo{year}{2014}), \eprint{1410.2242}.

\bibitem[{\citenamefont{{Doro} et~al.}(2013)\citenamefont{{Doro}, {Conrad},
  {Emmanoulopoulos}, {S{\`a}nchez-Conde}, {Barrio}, {Birsin}, {Bolmont},
  {Brun}, {Colafrancesco}, {Connell} et~al.}}]{2013APh....43..189D}
\bibinfo{author}{\bibfnamefont{M.}~\bibnamefont{{Doro}}},
  \bibinfo{author}{\bibfnamefont{J.}~\bibnamefont{{Conrad}}},
  \bibinfo{author}{\bibfnamefont{D.}~\bibnamefont{{Emmanoulopoulos}}},
  \bibinfo{author}{\bibfnamefont{M.~A.} \bibnamefont{{S{\`a}nchez-Conde}}},
  \bibinfo{author}{\bibfnamefont{J.~A.} \bibnamefont{{Barrio}}},
  \bibinfo{author}{\bibfnamefont{E.}~\bibnamefont{{Birsin}}},
  \bibinfo{author}{\bibfnamefont{J.}~\bibnamefont{{Bolmont}}},
  \bibinfo{author}{\bibfnamefont{P.}~\bibnamefont{{Brun}}},
  \bibinfo{author}{\bibfnamefont{S.}~\bibnamefont{{Colafrancesco}}},
  \bibinfo{author}{\bibfnamefont{S.~H.} \bibnamefont{{Connell}}},
  \bibnamefont{et~al.}, \bibinfo{journal}{Astroparticle Physics}
  \textbf{\bibinfo{volume}{43}}, \bibinfo{pages}{189} (\bibinfo{year}{2013}),
  \eprint{1208.5356}.

\bibitem[{\citenamefont{{Wood} et~al.}(2013)\citenamefont{{Wood}, {Buckley},
  {Digel}, {Funk}, {Nieto}, and {Sanchez-Conde}}}]{2013arXiv1305.0302W}
\bibinfo{author}{\bibfnamefont{M.}~\bibnamefont{{Wood}}},
  \bibinfo{author}{\bibfnamefont{J.}~\bibnamefont{{Buckley}}},
  \bibinfo{author}{\bibfnamefont{S.}~\bibnamefont{{Digel}}},
  \bibinfo{author}{\bibfnamefont{S.}~\bibnamefont{{Funk}}},
  \bibinfo{author}{\bibfnamefont{D.}~\bibnamefont{{Nieto}}}, \bibnamefont{and}
  \bibinfo{author}{\bibfnamefont{M.~A.} \bibnamefont{{Sanchez-Conde}}},
  \bibinfo{journal}{ArXiv e-prints}  (\bibinfo{year}{2013}),
  \eprint{1305.0302}.

\bibitem[{\citenamefont{{Weinstein} et~al.}(2013)\citenamefont{{Weinstein},
  {Dumm}, {Fortson}, and {Mukherjee}}}]{2013arXiv1305.0082W}
\bibinfo{author}{\bibfnamefont{A.}~\bibnamefont{{Weinstein}}},
  \bibinfo{author}{\bibfnamefont{J.}~\bibnamefont{{Dumm}}},
  \bibinfo{author}{\bibfnamefont{L.}~\bibnamefont{{Fortson}}},
  \bibnamefont{and}
  \bibinfo{author}{\bibfnamefont{R.}~\bibnamefont{{Mukherjee}}},
  \bibinfo{journal}{ArXiv e-prints}  (\bibinfo{year}{2013}),
  \eprint{1305.0082}.

\end{thebibliography}

\end{document}